\documentclass[a4paper,11pt]{article}

\usepackage[T1]{fontenc}
\usepackage[utf8]{inputenc}
\usepackage[english]{babel}
\usepackage[centertags]{amsmath}
\usepackage{amssymb,amsfonts,amsthm,mathrsfs,bm}
\usepackage[mathcal]{euscript}
\usepackage{graphicx}
\usepackage{xcolor}
\usepackage{braket}
\usepackage{slashed}
\usepackage{jheppub}

\newcommand{\be}{\begin{equation}}
\newcommand{\ee}{\end{equation}}
\newcommand{\bea}{\begin{eqnarray}}
\newcommand{\eea}{\end{eqnarray}}

\newcommand \tr {\mbox{{\bf Tr}}}
\def\Tr{\mathop{\rm Tr}}
\newcommand{\diag}{\mathop{\rm diag}}

\newcommand{\mC}{\mathscr C}

\newcommand{\cO}{{\cal O}}

\begin{document}

\begin{flushright}
INR-TH-2025-005
\end{flushright}

\title{\LARGE \bf Disparity in sound speeds: implications for elastic unitarity and the effective potential in quantum field theory}

\author{Dmitry S. Ageev$^{a,b}$ and Yulia A. Ageeva$^{c,d}$}

\affiliation{
$^{a}$Steklov Mathematical Institute of Russian Academy of Sciences,
Gubkin str. 8, 119991 Moscow, Russia\\
$^{b}$Institute for Theoretical and Mathematical Physics,
M.V. Lomonosov Moscow State University, 119991 Moscow, Russia\\
$^{c}$Institute for Nuclear Research of the Russian Academy of Sciences,
60th October Anniversary Prospect, 7a, 117312 Moscow, Russia\\
$^{d}$Department of Particle Physics and Cosmology, Physics Faculty, M.V. Lomonosov Moscow State University, 119991 Moscow, Russia
}

\emailAdd{ageev@mi-ras.ru, ageeva@inr.ac.ru}

\abstract{
We study interacting scalar field theories in which different fields propagate with inequivalent spatial kinetic tensors, corresponding to different sound speeds in different directions. We derive the exact elastic two-body unitarity relation and show that the phase space defines a positive kernel on the sphere, so that the scattering amplitude acts as an operator in angular-momentum space. The corresponding unitarity bounds constrain the eigenvalues of the phase-space-rescaled amplitude. In the weak-anisotropy regime, we obtain the leading correction explicitly and show that it induces $s-d$ mixing.

For a two-scalar quartic model, we verify the anisotropic optical theorem at one loop and derive coupled channel elastic unitarity bounds. We also compute the local one-loop effective potential and analyze the corresponding one-loop renormalization-group structure. In the classically scale-invariant limit, the Gildener-Weinberg flat direction is unchanged, whereas anisotropy modifies the radiatively generated scalon mass. In the isotropic but unequal-velocity limit, several results become analytic and the RG flow exhibits an additional invariant ray.
}

\maketitle
\tableofcontents

\section{Introduction}

In relativistic quantum field theory, exact Lorentz symmetry ties all excitations to a common light cone, and many standard results in perturbation theory and S-matrix theory exploit this shared kinematics.  In effective descriptions, however, it is common for different sectors to propagate with different characteristic speeds or, more generally, with inequivalent positive spatial quadratic forms.  This happens in cosmological scalar(-tensor) effective theories (where a ``sound speed'' governs fluctuations) \cite{Cheung:2007st,Armendariz-Picon:1999hyi,Garriga:1999vw,Alishahiha:2004eh,Kobayashi:2010cm,Kobayashi:2011nu,Creminelli:2010ba,Hinterbichler:2012fr,Pirtskhalava:2014esa,Ijjas:2016tpn,Kolevatov:2017voe,Mironov:2018oec,Ageeva:2022asq,Ageeva:2022byg,Ageeva:2023nwf,Deffayet:2010qz,Dobre:2017pnt,Babichev:2018twg,GilChoi:2025hbs}; it is also can be interesting  to consider the different spatial components of
sound speed in the models of domain walls (see, e.g.,
Refs.\cite{Dankovsky:2024zvs,Dankovsky:2024ipq,Babichev:2025stm} and references therein); it is supposed that the existence/taking
into account the different spatial components of speeds will crucially affect the evolution of domain
wall itself. Other setups are, for instance, condensed matter studies \cite{Ambegaokar,Ticknor,Blaschke,Belich:2002vd,Avila:2019xdn}, and/or the studies of astrophysical magnetic fields \cite{Arth:2014jea,Rappaz:2023rqd} and many others. From the viewpoint of microscopic QFT, these phenomena can be viewed as controlled departures from Lorentz invariance, in the same broad class as Lorentz-violating EFT frameworks \cite{Colladay:1998fq,Horava:2009uw,Myers:2003fd,Belich:2002vd,Avila:2019xdn,Rubtsov:2012kb,Satunin:2017wmk,Carvalho:2017stg,Altschul:2012xu,Altschul:2014gqa,Altschul:2022isc,Colladay:1996iz,Kostelecky:2001jc,Altschul:2023vnf}.  While Lorentz violation is often discussed in settings with modified dispersion relations or higher spatial derivatives (for instance in Ho\v{r}ava--Lifshitz-type constructions \cite{Horava:2009uw,Blas:2009qj,Barvinsky:2015kil,Barvinsky:2021ubv}), the kinematic input we consider is deliberately minimal: time derivatives remain canonical, stability is ensured by positive definiteness of the spatial tensors, and the theory remains power-counting renormalizable in four dimensions.  In this restricted but robust framework, one can cleanly isolate what is genuinely new already at the level of ordinary $2\to2$ scattering and one-loop vacuum functionals.

Concretely, we study multiplets of real scalars whose free quadratic action assigns to each field $\phi_i$ its own symmetric positive definite matrix $\mC_i$ multiplying spatial gradients.  A common linear redefinition of spatial coordinates acts on \textit{all} $\mC_i$ by the same congruence transformation.  Therefore, only \textit{relative} anisotropy is physical: for two fields one may choose a basis in which $\mC_1$ becomes the identity and $\mC_2$ becomes diagonal with eigenvalues $(v_x^2,v_y^2,v_z^2)$, but it is more invariant to package the mismatch into the relative matrix $\mC_1^{-1/2}\mC_2\mC_1^{-1/2}$.  Throughout the paper we keep a basis-covariant notation precisely because the same kinematic data will reappear in several guises: as directional quadratic forms in the two-body Jacobian, as harmonic moments of the phase-space kernel, and as geometric weights in the ultraviolet sector of the effective potential.  In this language the phrase ``disparity in sound speeds'' is shorthand for the existence of inequivalent propagation cones encoded in the collection $\{\mC_i\}$, not merely for a single isotropic velocity mismatch.

Once different excitations carry inequivalent spatial kinetic tensors, two features that are essentially automatic in the Lorentz-invariant case cease to be trivial.
First, the \textit{two-body phase space} at fixed center-of-mass energy is no longer a one-dimensional function of $E$ alone: the on-shell Jacobian depends on the direction of the relative momentum, so the available phase space becomes a \textit{positive kernel on the sphere}.  As a consequence, the familiar partial-wave unitarity argument \cite{Lee:1977yc,Lee:1977eg,Chanowitz:1985hj,Oller:2019opk,oller.190503.1,Lacour:2009ej,Gulmez:2016scm} must be reformulated: at fixed energy the $2\to2$ amplitude is generically an integral kernel $M(E;\hat{\mathbf n}',\hat{\mathbf n})$, and the unitarity relation becomes an operator equation in the combined channel in angular-momentum space rather than a set of decoupled scalar inequalities for $a_\ell(E)$.  Even when the microscopic interaction is a momentum-independent contact term (so that the \textit{tree} amplitude is angle independent), anisotropy re-enters through the on-shell measure once intermediate states are placed on shell.  In this sense the elastic unitarity problem is governed by a positive angular operator built from the phase-space weight. In the pure contact quartic model analyzed below, the explicit perturbative check probes only the $s$-wave component of this operator, whereas genuine higher-partial-wave amplitude mixing requires more general dynamical input. One of our goals is to make this statement exact and to formulate unitarity directly in the language in which it is naturally true. 

Second, the \textit{one-loop effective potential} in multi-field theories is usually organized by diagonalizing a momentum-independent mass matrix, leading (in simple cases) to Coleman--Weinberg-type expressions \cite{Coleman:1973jx,Martin:2001vx}.   With inequivalent spatial tensors $\mC_i$, the fluctuation operator cannot be diagonalized by a momentum-independent field rotation: the mixing angle is momentum dependent, and the determinant develops genuinely mixed structures that retain detailed memory of the relative anisotropy.  Nevertheless, the \textit{local} renormalizable ultraviolet sector can still be extracted systematically by expanding the determinant in background insertions. The renormalization in the presence of
spatial anisotropy (in the spirit of our sound speed disparity) in quantum field theory has been previously considered in \cite{Arefeva:1994qr,Periwal:1995wm}.  In our expansion the anisotropy organizes itself into distinct geometric weights: diagonal contractions are governed by $\Pi_i^{-1}=1/\sqrt{\det\mC_i}$, while mixed contractions are controlled by an interpolating invariant $\mathcal J_{12}=\int_0^1 dx\,(\det[x\mC_1+(1-x)\mC_2])^{-1/2}$.  This split is closely tied to general lessons from Lorentz-violating renormalization: the Lorentz-violating tensors typically enter counterterms and beta functions in a structured, operator-valued way rather than as a single overall prefactor. In particular, renormalization in Lorentz-violating scalar and Yukawa theories has been analyzed in a series of works by Altschul and collaborators \cite{Altschul:2006Yukawa,Ferrero:2011RenLV,Altschul:2012ScalarPot,Altschul:2014gqa}, and recent studies have extended related techniques to scalar gauge theories \cite{Altschul:2022ScalarQED,Altschul:2023ScalarQCD}.  The present model provides a setting in which these ideas can be seen in a particularly transparent way: in the pure quartic theory there is no one-loop wavefunction renormalization in the unbroken phase, so the running of the $\mC_i$ themselves vanishes at this order, and one can cleanly track how the static tensors enter both scattering phase space and the local effective action.

A further theme, which becomes explicit in our results, is that the same anisotropic data controls both infrared and ultraviolet aspects of the theory, but in \textit{different} combinations.  The elastic problem is governed by the directional phase-space kernel $g_\beta(E,\hat{\mathbf n})$ and its harmonic moments; the local ultraviolet sector is governed by $\Pi_i^{-1}$ and $\mathcal J_{12}$, which may be viewed as geometric averages over the same family of quadratic forms.  In particular limits these structures simplify.  When the tensors are isotropic but unequal, $\mC_i=u_i^2 I$, the angular dependence disappears while the disparity of speeds remains; in that case both the unitarity weights and the mixed logarithmic kernel become analytic functions of $(u_1,u_2)$ and one can expose additional RG structure (including an accidental invariant ray).  Conversely, near isotropy the leading nontrivial harmonic content is quadrupolar, so the first genuinely new partial-wave effect is a controlled $s$--$d$ mixing.

The purpose of this paper is to develop a unified and basis-covariant treatment of these issues for scalar theories with arbitrary positive spatial kinetic matrices $\mC_i$.
Our first goal is to derive the \textit{exact} two-body elastic unitarity relation in the presence of relative anisotropy and unequal propagation cones.  The key object is a positive phase-space kernel $g_\beta(E,\hat{\mathbf n})$ on $S^2$ and its harmonic-space image $G(E)$, which controls both the optical theorem and the correct perturbative unitarity bounds through the eigenvalues of the rescaled operator $G^{1/2}aG^{1/2}$.  Our second goal is to follow the same anisotropic data through a concrete renormalizable model: the two-scalar quartic theory.  In this setting we (i) check the anisotropic optical theorem explicitly at one loop, (ii) derive sharp perturbative elastic bounds, and (iii) compute the local one-loop effective potential and the corresponding one-loop beta functions.  We also discuss the classically scale-invariant limit in the spirit of Gildener--Weinberg \cite{Gildener:1976ih} and present a multiscale RG improvement adapted to the separation between diagonal and genuinely mixed logarithms.  The scope of the present paper is deliberately focused: we treat the exact two-body sector and the local one-loop effective potential in a theory with canonical time derivatives and renormalizable interactions.  This focus keeps the geometry of the problem fully explicit and prepares the ground for extensions to derivative interactions, larger multiplets, and settings in which kinetic tensors themselves run.

A more general one-loop renormalization framework for anisotropic two-scalar theories with the most general two-derivative Lorentz-violating quadratic form, including direction-dependent kinetic mixing and general interactions including cubic ones, has recently been developed in Ref.~\cite{Ageev:2025bwc}. In the present work we focus instead on the exact two-body unitarity problem and on the local one-loop effective potential in the simpler setting without kinetic mixing and only quartic interactions.

The paper is organized as follows.
Section~\ref{sec:unitarity} formulates the exact two-body kinematics with general $\mC_i$, derives the directional Jacobian and phase-space kernel, and establishes the exact angular-operator form of elastic unitarity.
Section~\ref{sec:quartic_model} applies this framework to the quartic two-scalar model, verifies the one-loop optical theorem, and extracts the coupled-channel eigenvalue bounds.
Section~\ref{sec:weakk} analyzes weak anisotropy (showing explicitly the leading $s$--$d$ mixing).
Section~\ref{sec:effpot} derives the local one-loop effective potential and the one-loop renormalization data, while section~\ref{sec:msrg} discusses the classically scale-invariant limit and multiscale RG improvement.
Section~\ref{sec:isotropic} gives the analytically tractable isotropic-but-unequal-velocity specialization, and section~\ref{sec:broken} discusses the broken-phase dispersion relations in the local-potential approximation.

\section{Anisotropic unitarity relation}
\label{sec:unitarity}
We begin with a set of real scalars $\phi_i$ whose quadratic action is
\begin{align}
S_{\rm free}
=
\sum_i \int d^4x\,
\left[
\frac12 \dot\phi_i^2
-\frac12 \partial_a\phi_i\,\mC_i^{ab}\,\partial_b\phi_i
-\frac12 m_i^2\phi_i^2
\right],
\label{eq:free_action_main}
\end{align}
with $\mC_i^{ab}=\mC_i^{ba}$ real, symmetric, and positive definite. The dispersion relation reads
\begin{align}
E_i(\mathbf p)^2 = m_i^2+\mathbf p^T\mC_i\mathbf p.
\label{eq:dispersion_main}
\end{align}

A common linear change of spatial coordinates acts on every kinetic matrix by the same transformation,
\begin{align}
\mC_i\longrightarrow \mC_i' = B^{-1}\mC_i B^{-T}.
\label{eq:congruence_main}
\end{align}
Therefore for two fields one may always choose a basis in which one of the matrices is diagonal
\begin{align}
\mC_1 = I,
\qquad
\mC_2 = \diag(v_x^2,v_y^2,v_z^2),
\label{eq:canonical_twofield_main}
\end{align}
so the invariant anisotropy is encoded in the spectrum of the relative matrix
\begin{align}
A\equiv \mC_1^{-1/2}\mC_2\mC_1^{-1/2}.
\label{eq:relative_matrix_main}
\end{align}
We nevertheless keep the basis-independent notation because it extends more naturally to larger multiplets and more general setups.

For a two-body channel, i.e. two distinguishable particles in a state the \textit{center-of-mass} (CM) kinematics is
\begin{align}
\mathbf p_1=\mathbf p,\qquad \mathbf p_2=-\mathbf p,\qquad \mathbf p=p\,\hat{\mathbf n},
\qquad p\ge 0,\qquad \hat{\mathbf n}^2=1.
\label{eq:com_kinematics_main}
\end{align}
It is convenient to define
\begin{align}
\omega_{1\beta}(\hat{\mathbf n}) \equiv \hat{\mathbf n}^T\mC_i\hat{\mathbf n},
\qquad
\omega_{2\beta}(\hat{\mathbf n}) \equiv \hat{\mathbf n}^T\mC_j\hat{\mathbf n},
\label{eq:omega_defs_main}
\end{align}
where $\beta\equiv  (\phi_i,\phi_j) \equiv(i,j)$. Let us clarify the notations here: index $1$ relates to the first particle in the pair $\beta$ which itself is a type $i$ (so that we write index ``$i$'' in ``$\mC_i$'') and $2$ is for the second particle in the pair which in its turn is ``$j$'' type. 
Then the dispersion relations are expressed as
\begin{align}
E_{1\beta}(p,\hat{\mathbf n}) = \sqrt{m_i^2+p^2\omega_{1\beta}(\hat{\mathbf n})},
\qquad
E_{2\beta}(p,\hat{\mathbf n}) = \sqrt{m_j^2+p^2\omega_{2\beta}(\hat{\mathbf n})},
\label{eq:Ei_beta_main}
\end{align}
and the total center-of-mass energy is $E=E_{1\beta}+E_{2\beta}$. Since
\begin{align}
\frac{\partial E}{\partial p}
=
\frac{p\,\omega_{1\beta}}{E_{1\beta}}
+
\frac{p\,\omega_{2\beta}}{E_{2\beta}}
>0,
\qquad (p>0),
\label{eq:monotonicity_main}
\end{align}
the on-shell condition determines a unique momentum magnitude
\begin{align}
p=p_\beta(E,\hat{\mathbf n})
\label{eq:p_beta_def_main}
\end{align}
for each open channel and each direction. It is worth writing this root explicitly, because it is the first place where the relative anisotropy enters in a sharp and invariant way. To package the answer, we introduce two directional kinematic combinations,
\begin{align}
\Delta_\beta(\hat{\mathbf n})
&\equiv
\omega_{1\beta}(\hat{\mathbf n})-\omega_{2\beta}(\hat{\mathbf n}),
\nonumber\\
\Xi_\beta(E,\hat{\mathbf n})
&\equiv
\big(\omega_{1\beta}(\hat{\mathbf n})+\omega_{2\beta}(\hat{\mathbf n})\big)E^2
-
\Delta_\beta(\hat{\mathbf n})(m_i^2-m_j^2),
\label{eq:DeltaXi_main}
\end{align}
together with the polynomial
\begin{align}
\mathcal{P}(x,y,z)\equiv x^2+y^2+z^2-2xy-2xz-2yz.
\label{eq:Kallen_main}
\end{align}
The combination $\Delta_\beta$ measures the directional mismatch between the two propagation cones, while $\Xi_\beta$ is the corresponding energy-weighted combination. Starting from
\begin{align}
E=\sqrt{m_i^2+p^2\omega_{1\beta}}+\sqrt{m_j^2+p^2\omega_{2\beta}},
\label{eq:E_constraint_app}
\end{align}
and squaring twice, one obtains a quadratic equation for $p^2$,
\begin{align}
\Delta_\beta(\hat{\mathbf n})^2\,p^4
-
2\Xi_\beta(E,\hat{\mathbf n})\,p^2
+
\mathcal{P}(E^2,m_i^2,m_j^2)
=0.
\label{eq:quadratic_p_app}
\end{align}
The physical branch is
\begin{align}
p_\beta(E,\hat{\mathbf n})^2
=
\frac{
\Xi_\beta(E,\hat{\mathbf n})
-
\sqrt{
\Xi_\beta(E,\hat{\mathbf n})^2
-
\Delta_\beta(\hat{\mathbf n})^2
\mathcal{P}(E^2,m_i^2,m_j^2)
}
}
{\Delta_\beta(\hat{\mathbf n})^2},
\label{eq:pbeta_explicit_main}
\end{align}
This solution  vanishes at threshold and matches continuously onto the isotropic limit. At $\Delta_\beta(\hat{\mathbf{n}})=0$ it reduces to
\begin{align}
p_\beta(E,\hat{\mathbf n})^2
=
\frac{\mathcal{P}(E^2,m_i^2,m_j^2)}
{4E^2\,\omega_\beta(\hat{\mathbf n})}
\qquad
\text{if }
\omega_{1\beta}=\omega_{2\beta}\equiv \omega_\beta.
\label{eq:p_smoothlimit_app}
\end{align}
This form is useful for two separate reasons. First, it shows explicitly that anisotropy enters only through the directional quadratic forms $\omega_{1\beta}$ and $\omega_{2\beta}$, rather than through any less transparent combination of matrix entries. Second, it makes the isotropic limit manifest.

Once $p_\beta(E,\hat{\mathbf n})$ is known, the individual on-shell energies are derived as directional functions of the total energy,
\begin{align}
E_{1\beta}(E,\hat{\mathbf n})
&=
\frac{
E^2+m_i^2-m_j^2+p_\beta(E,\hat{\mathbf n})^2\Delta_\beta(\hat{\mathbf n})
}{2E},
\nonumber\\
E_{2\beta}(E,\hat{\mathbf n})
&=
\frac{
E^2+m_j^2-m_i^2-p_\beta(E,\hat{\mathbf n})^2\Delta_\beta(\hat{\mathbf n})
}{2E}.
\label{eq:E12_closed_app}
\end{align}
Expressing the momentum-space volume element in spherical coordinates $d^3 p=p^2 d p d \Omega$, and then changing variables from the radial momentum $p$ to the total energy $E$, we obtain
\begin{align}
p^2\,dp\,d\Omega
=
dE\,d\Omega\,
\frac{p\,E_{1\beta}E_{2\beta}}
{E_{2\beta}\omega_{1\beta}(\hat{\mathbf n})+E_{1\beta}\omega_{2\beta}(\hat{\mathbf n})}.
\label{eq:phase_space_jac_main}
\end{align}
This follows from
\begin{align}
\frac{\partial E}{\partial p}
=
\frac{p\big(E_{2\beta}\omega_{1\beta}+E_{1\beta}\omega_{2\beta}\big)}{E_{1\beta}E_{2\beta}},
\label{eq:dEdp_app}
\end{align}
which is positive on every open channel. The denominator in \eqref{eq:phase_space_jac_main} is therefore  just the radial derivative of the total center-of-mass energy rewritten in a symmetric form and we will call it the \textit{phase-space kernel}. For distinguishable particles it is given by
\begin{align}
g_\beta(E,\hat{\mathbf n})
=
\frac{2p_\beta(E,\hat{\mathbf n})}
{E_{2\beta}(E,\hat{\mathbf n})\omega_{1\beta}(\hat{\mathbf n})
+
E_{1\beta}(E,\hat{\mathbf n})\omega_{2\beta}(\hat{\mathbf n})},
\label{eq:g_dist_main}
\end{align}
while for an identical particles in a pair $\beta=(i,i)$,
\begin{align}
g_i(E,\hat{\mathbf n})
=
\frac{p_i(E,\hat{\mathbf n})}{2E_i(E,\hat{\mathbf n})\,\omega_i(\hat{\mathbf n})}.
\label{eq:g_identical_main}
\end{align}
These kernels summarize the anisotropic two-body phase space. Every later appearance of anisotropy in the elastic problem will be mediated by $g_\beta(E,\hat{\mathbf n})$ or by the angular operator built from it. In particular, once $g_\beta$ acquires directional dependence, the angular variables are promoted from passive kinematic labels to dynamical data that enter the normalization, the projection procedure, and the structure of the unitarity condition.

Let us next turn to the terms of the total energy and the direction $\hat{\mathbf{n}}$, so that the three-dimensional delta function can be rewritten as
\begin{align}
\delta^{(3)}(\mathbf p'-\mathbf p)
=
\frac{
E_{2\beta}\omega_{1\beta}(\hat{\mathbf n})
+
E_{1\beta}\omega_{2\beta}(\hat{\mathbf n})
}
{p\,E_{1\beta}E_{2\beta}}
\,\delta(E'-E)\,\delta^{(2)}(\hat{\mathbf n}'-\hat{\mathbf n}),
\label{eq:delta3_app}
\end{align}
where variables with prime are related to the final state (and with no prime is for initial one). Let us clarify the rest of the notations in \eqref{eq:delta3_app}.
Having rewritten the latter in terms of the total energy and the direction $\hat{\mathbf{n}}$, it is natural to introduce two-body states labeled by these variables. To this end we first remind, that we consider
the center-of-mass frame \eqref{eq:com_kinematics_main} for the system of two colliding particles. Since below we are going to use partial wave expansion, the first step to this formalism is to replace the variables $\vec{p}_1$,
$\vec{p}_2$  \eqref{eq:com_kinematics_main}
by total momentum $P^{\mu} \equiv p_1^\mu + p_2^\mu$, where we
bear in mind that in CM frame we can choose
$P^{\mu} = (E, \mathbf{0})$ with $E= E_1 + E_2$. Throughout this section the kinematic functions $p_\beta(E, \hat{\mathbf{n}}), E_{1 \beta}(E, \hat{\mathbf{n}}), E_{2 \beta}(E, \hat{\mathbf{n}}), g_\beta(E, \hat{\mathbf{n}})$, and the angular normalization matrices introduced below are evaluated in the center-of-mass frame with $P^\mu=(E, \mathbf{0})$. We nevertheless retain the standard plane-wave label $P^\mu$ in the state normalization and completeness relation. Let us first consider a channel $\beta=(i, j)$ with $i \neq j$, so that the particles are distinguishable and no symmetrization issue arises. Restricting the standard two-particle plane-wave state to the center-of-mass slice, we may represent it as
\begin{align}
|\psi,P,\hat{\mathbf n},\beta\rangle
\equiv
\sqrt{2E_{1\beta}}\sqrt{2E_{2\beta}}\,
a_i^\dagger(\mathbf p)\,a_j^\dagger(-\mathbf p)\,|0\rangle,
\qquad
\beta=(i,j),
\qquad
i\neq j,
\label{eq:psi_dist_app}
\end{align}
where a notation $\psi$ is a shorthand for the pair of momenta,
$\psi \equiv \{ \vec{p}_1 , \vec{p}_2\}$. This state   satisfies
\begin{align}
\langle \psi',P',\hat{\mathbf n}',\beta'|\psi,P,\hat{\mathbf n},\beta\rangle
=\nonumber&\\=
(2\pi)^6
\frac{
4\big(
E_{2\beta}\omega_{1\beta}(\hat{\mathbf n})
+
E_{1\beta}\omega_{2\beta}(\hat{\mathbf n})
\big)
}{p_\beta(E,\hat{\mathbf n})}
\,&
\delta^{(4)}( P'- P)\,
\delta^{(2)}(\hat{\mathbf n}'-\hat{\mathbf n})\,
\delta_{\beta'\beta},
\label{eq:psi_norm_dist_app}
\end{align}
where we recall that
$ P  \equiv \sum  p_{\text{in}}$ and
$ P'  =\sum  p_{\text{out}}$ are total 4-momenta of the
initial and final state, respectively.  Here the factor $\delta^{(4)}\left(P^{\prime}-P\right)$  is kept in the standard plane-wave normalization. On the CM slice $P^{\prime}=\left(E^{\prime}, \mathbf{0}\right), P=(E, \mathbf{0})$, so that actually we have $\delta\left(E^{\prime}-E\right) \delta^{(3)}(\mathbf{0})$.

The identical-particle channel, $\beta=(i, i)$, requires a small but important modification. In addition to the direct contraction, one must include the exchange contraction, reflecting the symmetry under interchange of the two particles. As a result, the kinematics simplifies to
\begin{align}
\omega_\beta(\hat{\mathbf n}) \equiv \hat{\mathbf n}^T\mC_i\hat{\mathbf n},
\qquad
E_\beta = \frac{E}{2},
\qquad
p_\beta(E,\hat{\mathbf n}) = \frac{\sqrt{E^2/4-m_i^2}}{\sqrt{\omega_\beta(\hat{\mathbf n})}},
\label{eq:identical_kin_app}
\end{align}
and
\begin{align}
\langle \psi',P',\hat{\mathbf n}',\beta'|\psi,P,\hat{\mathbf n},\beta\rangle
=\nonumber&\\=
(2\pi)^6
\frac{
8\,E_\beta\,\omega_\beta(\hat{\mathbf n})
}{
p_\beta(E,\hat{\mathbf n})
}
\,&
\delta^{(4)}( P'- P)
\Big[
\delta^{(2)}(\hat{\mathbf n}'-\hat{\mathbf n})
+
\delta^{(2)}(\hat{\mathbf n}'+\hat{\mathbf n})
\Big]
\delta_{\beta'\beta}.
\label{eq:psi_norm_identical_app}
\end{align}

\subsection{Angular-space formulation of unitarity}

Because the phase space now depends on $\hat{\mathbf n}$, the scattering amplitude is no longer a function only of the relative angle between the incoming and outgoing momenta. Instead, at fixed center-of-mass energy it depends separately on the incoming and outgoing directions. 
Thus, for two-body asymptotic states with total four-momentum $P$ in CM frame, we define 
\begin{align}
\langle \psi,P',\hat{\mathbf n}',\beta'|\,T\,|\psi,P,\hat{\mathbf n},\beta\rangle
=
(2\pi)^4\delta^{(4)}( P'- P)\,
M_{\beta'\beta}(E;\hat{\mathbf n}',\hat{\mathbf n}),
\label{eq:T_kernel_main}
\end{align}
and we remind that the kinematics is evaluated in the CM frame:
\begin{align}
\left.
\big\langle \psi,P',\hat{\mathbf n}',\beta' \big|\,T\,\big| \psi,P,\hat{\mathbf n},\beta \big\rangle
\right|_{P'=(E',\mathbf 0),\,P=(E,\mathbf 0)}
\equiv
\big\langle E',\hat{\mathbf n}',\beta' \big|\,T\,\big| E,\hat{\mathbf n},\beta \big\rangle_{\mathrm{CM}}.
\end{align}
Thus $M_{\beta'\beta}(E;\hat{\mathbf n}',\hat{\mathbf n})$ is the CM $2\to2$ scattering matrix element between an incoming channel $\beta$ moving in direction $\hat{\mathbf n}$ and an outgoing channel $\beta'$ moving in direction $\hat{\mathbf n}'$. In the rotational invariant case this element collapses to a function of the single variable $\hat{\mathbf n}'\!\cdot\!\hat{\mathbf n}$, but in the present anisotropic problem that simplification is lost.

Next we introduce the angular basis
\begin{align}
|L,P,\beta\rangle
=
\frac{1}{\sqrt{4\pi}}
\int d\Omega_{\hat{\mathbf n}}\,
Y_L(\hat{\mathbf n})\,
|\psi,P,\hat{\mathbf n},\beta\rangle,
\label{eq:angular_basis_main}
\end{align}
where $L\equiv (l,m)$,
and define the angular-space matrix elements of the scattering operator by
\begin{align}
\langle L',P',\beta'|\,T\,|L,P,\beta\rangle
=
(2\pi)^4\delta^{(4)}( P'- P)\,
16\pi\,a_{L'L,\beta'\beta}(E).
\label{eq:T_a_relation_main}
\end{align}
Thus, inserting \eqref{eq:angular_basis_main} into \eqref{eq:T_kernel_main}, one obtains
\begin{align}
a_{L'L,\beta'\beta}(E)
=
\frac{1}{64\pi^2}
\int d\Omega' d\Omega\,
Y_{L'}^*(\hat{\mathbf n}')\,
M_{\beta'\beta}(E;\hat{\mathbf n}',\hat{\mathbf n})\,
Y_L(\hat{\mathbf n}).
\label{eq:a_def_main}
\end{align}
The quantity $a_{L'L,\beta'\beta}(E)$ is therefore the anisotropic analogue of the usual partial-wave amplitude: it is the matrix of the two-body scattering operator in the spherical-harmonic basis. In the rotationally invariant limit it becomes diagonal in $(l,m)$ and reduces to the ordinary partial waves,
\begin{align}
a_{L'L,\beta'\beta}(E)
=
\delta_{l'l}\delta_{m'm}\,a_{l,\beta'\beta}(E).
\label{eq:isotropic_a_diag_main}
\end{align}
In the generic anisotropic problem, however, different angular momenta mix, so $a(E)$ must be treated as an operator in the combined channel and angular-momentum space.
Hence, we also need \textit{angular normalization matrix}, namely the Gram matrix of two-body states at fixed total energy. For a distinguishable channel the corresponding scalar product is
\begin{align}
\langle L',P',\beta'|L,P,\beta\rangle
=
(2\pi)^4\delta^{(4)}( P'- P)\,\delta_{\beta'\beta}\,
\mathcal N^\beta_{L'L}(E),
\label{eq:N_def_app}
\end{align}
with
\begin{align}
\mathcal N^\beta_{L'L}(E)
=
4\pi
\int d\Omega\,
Y_{L'}^*(\hat{\mathbf n})\,
\frac{
E_{2\beta}\omega_{1\beta}(\hat{\mathbf n})
+
E_{1\beta}\omega_{2\beta}(\hat{\mathbf n})
}{
p_\beta(E,\hat{\mathbf n})
}
\,Y_L(\hat{\mathbf n}).
\label{eq:Nmatrix_main}
\end{align}
For identical particles in the pair one obtains the familiar even-parity projection together with
\begin{align}
\mathcal N^{\beta,{\rm id}}_{L'L}(E)
=
16\pi
\int d\Omega\,
Y_{L'}^*(\hat{\mathbf n})\,
\frac{E_\beta\,\omega_\beta(\hat{\mathbf n})}{p_\beta(E,\hat{\mathbf n})}\,
Y_L(\hat{\mathbf n}),
\qquad
l,l' \text{ even}.
\label{eq:Nmatrix_id_app}
\end{align}
These matrices are trivial only when the weight in the integral is constant, that is, only when the underlying kinematics is rotationally symmetric. In the anisotropic problem they are the first signal that different harmonics are coupled already before any dynamics is inserted.

The corresponding projector onto the two-body sector is
\begin{align}
1_{2{\rm body}}
=
\sum_\beta
\int\frac{dE\,d^3\mathbf P}{(2\pi)^4}
\sum_{L,L'}
|L,P,\beta\rangle
\big[\mathcal N^\beta(E)^{-1}\big]_{LL'}
\langle L',P,\beta|.
\label{eq:projector_app}
\end{align}
For identical channels, $\mathcal{N}^\beta(E)$ in \eqref{eq:projector_app} is understood as $\mathcal{N}^{\beta, \text { id }}(E)$, and the sums over $L, L^{\prime}$ are restricted to even $l, l^{\prime}$. We also note, that in the chosen plane-wave normalization, the factor $\delta^{(4)}\left(P^{\prime}-P\right)$ is saturated in the standard way by the integration $\int d E d^3 \mathbf{P} /(2 \pi)^4$ in \eqref{eq:projector_app}, so no separate volume normalization is introduced to cancel mentioned earlier $\delta^{(3)}(\mathbf{0})$.
So, at this point the role of the angular normalization matrix becomes completely concrete: it is the object that converts the non-orthonormal harmonic basis of two-body states into the correct resolution of the identity at fixed energy.

A closely related quantity is the phase-space operator
\begin{align}
&G^\gamma_{MN}(E)
=
8\pi\big[\mathcal N^\gamma(E)^{-1}\big]_{MN}
=
\int d\Omega\,
Y_M^*(\hat{\mathbf n})\,g_\gamma(E,\hat{\mathbf n})\,Y_N(\hat{\mathbf n}), \label{eq:G_operator_main}
\\
&G(E)=\bigoplus_\gamma G^\gamma(E),\nonumber
\end{align}
where  $\bigoplus_\gamma$ denotes the direct sum over open two-body channels $\gamma$, so that $G(E)$ is block diagonal in channel space, with each block $G^\gamma(E)$ acting on the angular-momentum indices within the corresponding channel.
This operator is the angular-space image of the on-shell two-body measure. Whereas $\mathcal N^\beta$ tells us how angular states are normalized in the anisotropic kinematics, $G^\gamma$ expresses how the available phase space is distributed among those same angular components once an intermediate channel is placed on shell. In a rotationally invariant problem both objects would be diagonal and essentially trivial. Here, by contrast, their off-diagonal structure is precisely what keeps track of the angular mixing induced by anisotropic phase space.

Projecting the operator relation $T-T^\dagger=iT^\dagger T$ onto the two-body sector yields
\begin{align}
-\frac{i}{2}\big(a-a^\dagger\big)=a^\dagger\,G\,a+X(E),
\qquad
X(E)\ge 0,
\label{eq:unitarity_matrix_main}
\end{align}
where $X(E)$ is the positive semidefinite contribution of inelastic states. In the purely elastic regime, or below the first inelastic threshold it reduces to
\begin{align}
-\frac{i}{2}\big(a-a^\dagger\big)=a\,G\,a^\dagger.
\label{eq:elastic_unitarity_main}
\end{align}
This is the exact anisotropic analogue of the ordinary partial-wave relation.

It is useful to rescale the amplitude by the positive operator $G$,
\begin{align}
\widetilde a(E)\equiv G(E)^{1/2}a(E)G(E)^{1/2}.
\label{eq:rescaled_a_main}
\end{align}
Then elastic unitarity becomes
\begin{align}
-\frac{i}{2}\big(\widetilde a-\widetilde a^\dagger\big)
=
\widetilde a\,\widetilde a^\dagger
=
\widetilde a^\dagger \widetilde a,
\label{eq:rescaled_unitarity_main}
\end{align}
so $\widetilde a$ is normal and each of its eigenvalues obeys
\begin{align}
\Im\lambda_n = |\lambda_n|^2,
\qquad
|\Re\lambda_n|\le \frac12,
\label{eq:eigenvalue_bound_main}
\end{align}
which is the correct unitarity bound in the anisotropic case (see \cite{Grojean:2007zz,Ageeva:2022byg,DuasoPueyo:2024usw} and references therein for further discussion): it constrains the eigenvalues of the rescaled operator in the combined channel and angular-momentum space.

\subsection{Harmonic mixing, selection rules, and positivity}

It is also useful to display the angular mixing in fully explicit form, because this is the point at which the anisotropic phase-space measure stops being a mere weight and becomes an operator that couples different harmonics. To do so, we expand the phase-space kernel itself in spherical harmonics,
\begin{align}
g_\gamma(E,\hat{\mathbf n})
=
\sum_{J=0}^{\infty}\sum_{M=-J}^{J}
g^\gamma_{JM}(E)\,Y_{JM}(\hat{\mathbf n}),
\label{eq:g_expand_main}
\end{align}
where the coefficients are the harmonic moments of the kernel,
\begin{align}
g^\gamma_{JM}(E)
=
\int d\Omega\,
Y_{JM}^*(\hat{\mathbf n})\,g_\gamma(E,\hat{\mathbf n}).
\label{eq:gJM_def_main}
\end{align}
Substituting this expansion into the definition of the angular operator,
\begin{align}
G^\gamma_{l'm';lm}(E)
=
\int d\Omega\,
Y_{l'm'}^*(\hat{\mathbf n})\,
g_\gamma(E,\hat{\mathbf n})\,
Y_{lm}(\hat{\mathbf n}),
\label{eq:G_def_again_main}
\end{align}
one is led to the integral of three spherical harmonics. Using the standard Gaunt formula in terms of Wigner $3j$ symbols,
\begin{align}
&\int d\Omega\,
Y_{l'm'}^*(\hat{\mathbf n})\,Y_{JM}(\hat{\mathbf n})\,Y_{lm}(\hat{\mathbf n})\nonumber\\
&=
(-1)^{m'}
\sqrt{\frac{(2l'+1)(2J+1)(2l+1)}{4\pi}}
\begin{pmatrix}
l' & J & l\\
0 & 0 & 0
\end{pmatrix}
\begin{pmatrix}
l' & J & l\\
-m' & M & m
\end{pmatrix},
\label{eq:Gaunt_main}
\end{align}
we obtain
\begin{align}
&G^\gamma_{l'm';lm}(E)\nonumber\\
&=
\sum_{J,M}
g^\gamma_{JM}(E)\,
(-1)^{m'}
\sqrt{\frac{(2l'+1)(2J+1)(2l+1)}{4\pi}}
\begin{pmatrix}
l' & J & l\\
0 & 0 & 0
\end{pmatrix}
\begin{pmatrix}
l' & J & l\\
-m' & M & m
\end{pmatrix}.
\label{eq:G_3j_main}
\end{align}
The coefficients $g^\gamma_{JM}(E)$ encode the angular part of the two-body phase-space kernel itself: they are nonzero only to the extent that the available phase space depends on direction. The first Wigner $3j$ symbol controls which orbital angular momenta are allowed to couple, while the second keeps track of the corresponding $m$ quantum numbers. In this sense, the operator $G^\gamma(E)$ inherits its entire angular structure from the harmonic decomposition of $g_\gamma(E,\hat{\mathbf n})$.

The selection rules follow immediately from the properties of the $3j$ symbols:
\begin{align}
|l-J|\le l' \le l+J,
\qquad
m'=m+M,
\qquad
l'+J+l \ \text{even}.
\label{eq:G_selection_main}
\end{align}
The first condition is the familiar triangle inequality, the second expresses conservation of the $m$-quantum number, and the third is the parity constraint. Since the phase-space kernel is even under inversion of the direction,
\begin{align}
g_\gamma(E,-\hat{\mathbf n})=g_\gamma(E,\hat{\mathbf n}),
\label{eq:g_even_main}
\end{align}
its harmonic expansion contains only even values of $J$. It follows that anisotropy mixes only partial waves of the same parity. In particular, near isotropy the leading nontrivial contribution comes from the quadrupole sector $J=2$, so the first new coupling generated by the anisotropic phase-space operator is the mixing between the $s$- and $d$-waves (recalling that $s$-wave relates to $l=0$, while $d$-wave to $l=2$).

Finally, let us comment, that for any set of coefficients $c_{lm}$,
\begin{align}
\sum_{l'm',lm} c_{l'm'}^*\,G^\gamma_{l'm';lm}(E)\,c_{lm}
=
\int d\Omega\,
g_\gamma(E,\hat{\mathbf n})
\left|
\sum_{lm} c_{lm}Y_{lm}(\hat{\mathbf n})
\right|^2
\ge 0,
\label{eq:G_positive_main}
\end{align}
because the phase-space kernel is pointwise non-negative on every open channel. Thus $G^\gamma(E)$ is a positive semidefinite operator on angular space, as required by its role in the elastic unitarity relation.

\section{The quartic two-scalar model: optical theorem and unitarity bounds}
\label{sec:quartic_model}

We now turn from the general formalism to the simplest renormalizable setting in which all of its features can be displayed concretely, namely a two-scalar quartic theory with inequivalent spatial kinetic tensors. This model  exhibits the interplay between relative anisotropy, coupled-channel elastic unitarity, and the structure of the one-loop effective potential, while still remaining sufficiently simple. Its action is
\begin{align}
S
=
\int d^4x\,
\Bigg[
&\,
\frac12\dot\phi_1^2+\frac12\dot\phi_2^2
-\frac12\partial_a\phi_1\,\mC_1^{ab}\,\partial_b\phi_1
-\frac12\partial_a\phi_2\,\mC_2^{ab}\,\partial_b\phi_2
-\frac12 m_1^2\phi_1^2
-\frac12 m_2^2\phi_2^2
\nonumber\\[1mm]
&\,
-\frac{\lambda_1}{4!}\phi_1^4
-\frac{\lambda_2}{4!}\phi_2^4
-\frac{\lambda_3}{4}\phi_1^2\phi_2^2
\Bigg].
\label{eq:model_quartic_main}
\end{align}
Correspondingly, the elastic two-body problem involves three states:
\begin{align}
\alpha=(\phi_1,\phi_1),
\qquad
\beta=(\phi_1,\phi_2),
\qquad
\gamma=(\phi_2,\phi_2).
\label{eq:channel_basis_model_main}
\end{align}
This basis is convenient because it aligns directly with the coupled-channel structure induced by the quartic vertices.

At tree level the dynamics is particularly simple. Since all interactions are non-derivative, the corresponding $2\to2$ amplitudes do not depend on the scattering directions. In other words, the nontrivial angular structure discussed in the previous section is not yet generated by the vertices themselves; it enters only through the phase-space measure once intermediate states are placed on-shell. The tree-level amplitude matrix element is therefore angle independent and takes the form
\begin{align}
M_{\rm tree}
=
-
\begin{pmatrix}
\lambda_1 & 0 & \lambda_3\\
0 & \lambda_3 & 0\\
\lambda_3 & 0 & \lambda_2
\end{pmatrix},
\qquad
a^{\rm tree}_{00}(E)=\frac{M_{\rm tree}}{16\pi},
\label{eq:tree_matrix_main}
\end{align}
and index ``$00$'' comes from \eqref{eq:isotropic_a_diag_main}, i.e. here we have $L' = L = 0$. 
At this order, the amplitude lives entirely in the $Y_{00}$ sector: only the $s$-wave is populated, and all higher harmonics vanish. This makes the model an especially clean testing ground for the general framework, because the first nontrivial angular effects can then be traced unambiguously to anisotropic phase space rather than to any angular dependence already present in the interaction vertex.

\subsection{A direct check of the general anisotropic optical theorem}

The one-loop matrix element's part comes from the bubble diagram. This graph is the natural testing ground for the general formalism, because it forces all of the ingredients introduced above to work together: the unusual dispersion relations, the directional phase-space measure, and the identical-particle combinatorics. The one-loop absorptive part comes from the bubble diagram. For a transition $\alpha\to\beta$ through an intermediate channel $\gamma=(i,j)$, the loop integral is
\begin{align}
&iM^{(1)}_{\alpha\beta|\gamma}(E)\nonumber\\
&=
\eta_\gamma
M^{\rm tree}_{\alpha\gamma}\left[
\int \frac{d^4q}{(2\pi)^4}
\frac{1}{
\left(\frac{E}{2}-q^0\right)^2-\Omega_i(\mathbf q)^2+i0
}
\frac{1}{
\left(\frac{E}{2}+q^0\right)^2-\Omega_j(\mathbf q)^2+i0
}\right]M^{\rm tree*}_{\beta\gamma},
\label{eq:bubble_integral_main}
\end{align}
where $\eta_\gamma=1$ for a distinguishable intermediate pair and $\eta_\gamma=\tfrac12$ for an identical one. The two factors $
M^{\rm tree}_{\alpha\gamma},
$ and
$
M^{\rm tree*}_{\beta\gamma},
$
are the tree-level amplitudes for entering and leaving the intermediate channel, so the loop is organized exactly as one expects from unitarity: a tree process on the left, an on-shell propagation measure in the middle, and a tree process on the right. The integration variable $q^\mu=(q^0,\mathbf q)$ is the loop four-momentum, while the two propagators describe the intermediate particles carrying center-of-mass energies $E/2\mp q^0$ and opposite spatial momenta $\pm\mathbf q$. Their anisotropic dispersion laws are encoded in
\begin{align}
\Omega_i(\mathbf q)=\sqrt{m_i^2+\mathbf q^T\mC_i\mathbf q},
\qquad
\Omega_j(\mathbf q)=\sqrt{m_j^2+\mathbf q^T\mC_j\mathbf q},
\end{align}
so each propagator remembers the kinetic matrix of the corresponding field. The Feynman prescription $+i0$ fixes the pole locations in the usual causal way and, in particular, determines which poles pinch the contour when the intermediate state can go on shell.

It is useful to stress that the dependence on the relative anisotropy enters only through the two one-particle energies $\Omega_i(\mathbf q)$ and $\Omega_j(\mathbf q)$.

One may now proceed in two logically distinct but physically equivalent ways. The first is to do the $q^0$ contour integral and isolate the discontinuity of the physical $s$-channel pole. The second is to cut the graph directly and replace each propagator by its positive-energy on-shell delta function.

After closing the contour in the lower half-plane, the pole associated with $q^0=-E/2+\Omega_j-i0$ is the one that participates in the relevant pinching configuration. The remaining residue contributes only to the dispersive part. Thus the imaginary part is
\begin{align}
\Im M^{(1)}_{\alpha\beta|\gamma}(E)
=
\frac{
\eta_\gamma\,
M^{\rm tree}_{\alpha\gamma}
M^{\rm tree*}_{\beta\gamma}
}{32\pi^2}
\int d^3q\,
\frac{
\delta(E-\Omega_i(\mathbf q)-\Omega_j(\mathbf q))
}{
\Omega_i(\mathbf q)\Omega_j(\mathbf q)
}.
\label{eq:ImM_qspace_app}
\end{align}
Using the Jacobian derived in section~\ref{sec:unitarity} yields
\begin{align}
\Im M^{(1)}_{\alpha\beta|\gamma}(E)
=
\frac{
\eta_\gamma\,
M^{\rm tree}_{\alpha\gamma}
M^{\rm tree*}_{\beta\gamma}
}{32\pi^2}
\int d\Omega\,
\frac{
p_\gamma(E,\hat{\mathbf n})
}{
E_{2\gamma}(E,\hat{\mathbf n})\omega_{1\gamma}(\hat{\mathbf n})
+
E_{1\gamma}(E,\hat{\mathbf n})\omega_{2\gamma}(\hat{\mathbf n})
}.
\label{eq:ImM_angular_app}
\end{align}
For an identical intermediate channel $\gamma=(i,i)$ this simplifies to
\begin{align}
\Im M^{(1)}_{\alpha\beta|(i,i)}(E)
=
\frac{
M^{\rm tree}_{\alpha,(i,i)}
M^{\rm tree*}_{\beta,(i,i)}
}{64\pi^2}
\int d\Omega\,
\frac{p_i(E,\hat{\mathbf n})}{E\,\omega_i(\hat{\mathbf n})}.
\label{eq:ImM_identical_app}
\end{align}

The same result follows directly from the cut. Here the logic is even more transparent: one simply replaces each propagator by its positive-energy on-shell delta function, thereby forcing the two internal lines onto the physical intermediate state. The cut computation therefore makes the optical theorem visible in a single formula,
\begin{align}
&2\Im M^{(1)}_{\alpha\beta|\gamma}(E)\nonumber\\
&=
\eta_\gamma
M^{\rm tree}_{\alpha\gamma}
M^{\rm tree*}_{\beta\gamma}
\int \frac{d^4q}{(2\pi)^4}
(2\pi)\delta_+\!\left(\left(\frac{E}{2}-q^0\right)^2-\Omega_i^2\right)
(2\pi)\delta_+\!\left(\left(\frac{E}{2}+q^0\right)^2-\Omega_j^2\right),
\label{eq:cut_formula_app}
\end{align}
which reproduces \eqref{eq:ImM_qspace_app} after the $q^0$ integral.

For identical intermediate particles expression \eqref{eq:ImM_identical_app} we first write
\begin{align}
p_i(E,\hat{\mathbf n})=\frac{\sqrt{E^2/4-m_i^2}}{\sqrt{\omega_i(\hat{\mathbf n})}},
\label{eq:pi_ident_app}
\end{align}
then together with \eqref{eq:identical_kin_app}, we arrive to
\begin{align}
\Im M^{(1)}_{\alpha\beta|(i,i)}(E)
=
\frac{
M^{\rm tree}_{\alpha,(i,i)}
M^{\rm tree*}_{\beta,(i,i)}
}{16\pi}
\frac{\sqrt{E^2/4-m_i^2}}{E\,\sqrt{\det\mC_i}},
\label{eq:ImM_identical_final_app}
\end{align}
where we also use 
\begin{align}
\int d\Omega\,\omega_i(\hat{\mathbf n})^{-3/2}
=
\frac{4\pi}{\sqrt{\det\mC_i}},
\label{eq:ellipsoid_identity_app}
\end{align}
what brings the full directional dependence in this case, since the two propagators in the loop carry the same kinetic matrix. By contrast, the mixed factor retains genuine memory of the full relative anisotropy encoded in \eqref{eq:relative_matrix_main}. In that sense the mixed bubble is the sharpest probe of the mismatch between the two propagation cones.

Collecting the channel contributions leads to the compact matrix result
\begin{align}
\Im M^{(1)}_{\rm loop}(E)
=
\frac{1}{16\pi}\,
M_{\rm tree}\,D(E)\,M_{\rm tree},
\qquad
D(E)=\diag\!\big(d_1(E),d_{12}(E),d_2(E)\big),
\label{eq:ImM_loop_main}
\end{align}
where the identical-channel weights are
\begin{align}
d_1(E)=\frac{\sqrt{E^2/4-m_1^2}}{E\,\sqrt{\det\mC_1}},
\qquad
d_2(E)=\frac{\sqrt{E^2/4-m_2^2}}{E\,\sqrt{\det\mC_2}},
\label{eq:d1d2_main}
\end{align}
and the mixed weight is
\begin{align}
d_{12}(E)
=
\frac{1}{4\pi}
\int d\Omega\,
\frac{2p_{12}(E,\hat{\mathbf n})}
{E_2(E,\hat{\mathbf n})\omega_1(\hat{\mathbf n})+E_1(E,\hat{\mathbf n})\omega_2(\hat{\mathbf n})}.
\label{eq:d12_main}
\end{align}
Since $a^{\rm tree}_{00}=M_{\rm tree}/(16\pi)$, eq.~\eqref{eq:ImM_loop_main} is equivalently
\begin{align}
\Im a^{(1)}_{00}(E)
=
a^{\rm tree}_{00}(E)\,D(E)\,a^{\rm tree}_{00}(E),
\label{eq:Im_a00_main}
\end{align}
which is precisely the $s$-wave projection of the exact elastic two-body relation \eqref{eq:elastic_unitarity_main}. The explicit one-loop computation therefore furnishes a direct check of the general anisotropic optical theorem.

The two following asymptotic regimes are worth mentioning. Near an open threshold $E_{ij}^{\rm th}=m_i+m_j$, write $\delta E=E-E_{ij}^{\rm th}$. Expanding the exact center-of-mass relation at small $p$ gives
\begin{align}
E
=
m_i+m_j
+
\frac{p^2}{2}
\left(
\frac{\omega_i(\hat{\mathbf n})}{m_i}
+
\frac{\omega_j(\hat{\mathbf n})}{m_j}
\right)
+O(p^4),
\label{eq:threshold_expand_p_app}
\end{align}
hence
\begin{align}
p(E,\hat{\mathbf n})
=
\sqrt{
\frac{
2\delta E
}{
\omega_i(\hat{\mathbf n})/m_i+\omega_j(\hat{\mathbf n})/m_j
}
}
+
O\big((\delta E)^{3/2}\big),
\label{eq:p_threshold_app}
\end{align}
and
\begin{align}
g_{ij}(E,\hat{\mathbf n})
=
2\sqrt{2m_im_j}\,
\frac{
\sqrt{\delta E}
}{
\big(m_j\omega_i(\hat{\mathbf n})+m_i\omega_j(\hat{\mathbf n})\big)^{3/2}
}
+
O\big((\delta E)^{3/2}\big).
\label{eq:g_threshold_app}
\end{align}
Averaging over the sphere gives
\begin{align}
c_{ij}^{\rm anisotropic}
=
\frac{\sqrt{2m_im_j}}{2\pi}
\int d\Omega\,
\frac{1}{
\big(m_j\omega_i(\hat{\mathbf n})+m_i\omega_j(\hat{\mathbf n})\big)^{3/2}
},
\label{eq:cij_app}
\end{align}
and therefore
\begin{align}
d_{ij}(E)=c_{ij}^{\rm anisotropic}\,\sqrt{E-E_{ij}^{\rm th}}+O\!\big((E-E_{ij}^{\rm th})^{3/2}\big).
\label{eq:threshold_dij_main}
\end{align}
In other words, anisotropy does not change the universal threshold exponent; it changes the geometric prefactor that multiplies it. For an identical channel,
\begin{align}
d_i(E)
=
\frac{\sqrt{E-2m_i}}{2\sqrt{m_i}\,\sqrt{\det\mC_i}}
+O\!\big((E-2m_i)^{3/2}\big).
\label{eq:threshold_di_main}
\end{align}

At large energy, by contrast, the masses retreat to subleading order and the geometry of the propagation cones takes over. Writing $s_i(\hat{\mathbf n})\equiv \sqrt{\omega_i(\hat{\mathbf n})}$ and $s_j(\hat{\mathbf n})\equiv \sqrt{\omega_j(\hat{\mathbf n})}$, one finds
\begin{align}
p(E,\hat{\mathbf n})
=
\frac{E}{s_i(\hat{\mathbf n})+s_j(\hat{\mathbf n})}
-\frac{1}{2E}
\left(
\frac{m_i^2}{s_i(\hat{\mathbf n})}
+\frac{m_j^2}{s_j(\hat{\mathbf n})}
\right)
+O(E^{-3}),
\label{eq:p_uv_app}
\end{align}
and therefore
\begin{align}
g_{ij}(E,\hat{\mathbf n})
=
\frac{
2
}{
\sqrt{\omega_i(\hat{\mathbf n})}\sqrt{\omega_j(\hat{\mathbf n})}
\big(\sqrt{\omega_i(\hat{\mathbf n})}+\sqrt{\omega_j(\hat{\mathbf n})}\big)
}
+O(E^{-2}),
\label{eq:g_uv_app}
\end{align}
which leads to
\begin{align}
d_i(E)\longrightarrow \frac{1}{2\sqrt{\det\mC_i}},
\qquad
d_{ij}(E)\longrightarrow
\frac{1}{4\pi}
\int d\Omega\,
\frac{2}{\sqrt{\omega_i(\hat{\mathbf n})}\sqrt{\omega_j(\hat{\mathbf n})}\big(\sqrt{\omega_i(\hat{\mathbf n})}+\sqrt{\omega_j(\hat{\mathbf n})}\big)}.
\label{eq:UV_d_main}
\end{align}
For an identical channel one may even keep the subleading terms explicitly,
\begin{align}
d_i(E)
=
\frac{1}{2\sqrt{\det\mC_i}}
\sqrt{1-\frac{4m_i^2}{E^2}}
=
\frac{1}{2\sqrt{\det\mC_i}}
\left[
1-\frac{2m_i^2}{E^2}-\frac{2m_i^4}{E^4}+O(E^{-6})
\right].
\label{eq:di_uv_app}
\end{align}

A useful specialization is obtained by taking
\begin{align}
\mC_1=u_1^2 I,
\qquad
\mC_2=u_2^2 I,
\qquad
u_1\neq u_2,
\end{align}
so that the directional factors become constants,
\begin{align}
\omega_1(\hat{\mathbf n})=u_1^2,
\qquad
\omega_2(\hat{\mathbf n})=u_2^2.
\end{align}
The angular dependence of the cut bubble then disappears completely. In particular, the mixed phase-space weight
\begin{align}
d_{12}(E)
=
\frac{1}{4\pi}\int d\Omega\,
\frac{2p_{12}(E,\hat{\mathbf n})}
{E_2(E,\hat{\mathbf n})\omega_1(\hat{\mathbf n})+E_1(E,\hat{\mathbf n})\omega_2(\hat{\mathbf n})}
\end{align}
reduces to
\begin{align}
d_{12}(E)
=
\frac{2p_{12}(E)}
{E_2(E)\,u_1^2+E_1(E)\,u_2^2},
\label{eq:d12_iso_drop}
\end{align}
where
\begin{align}
E_1(E)
&=
\frac{E^2+m_1^2-m_2^2+(u_1^2-u_2^2)p_{12}(E)^2}{2E},
\nonumber\\
E_2(E)
&=
\frac{E^2+m_2^2-m_1^2-(u_1^2-u_2^2)p_{12}(E)^2}{2E},
\label{eq:E12_iso_drop}
\end{align}
and the center-of-mass momentum is the physical root of
\begin{align}
E=\sqrt{m_1^2+u_1^2p^2}+\sqrt{m_2^2+u_2^2p^2}.
\end{align}
Equivalently,
\begin{align}
p_{12}(E)^2
=
\frac{
\Xi(E)-\sqrt{\Xi(E)^2-(u_1^2-u_2^2)^2\mathcal{P}(E^2,m_1^2,m_2^2)}
}{
(u_1^2-u_2^2)^2
},
\label{eq:p12_iso_drop}
\end{align}
with
\begin{align}
\Xi(E)
=
(u_1^2+u_2^2)E^2-(u_1^2-u_2^2)(m_1^2-m_2^2).
\end{align}
For the identical channels one similarly finds
\begin{align}
d_1(E)=\frac{\sqrt{E^2/4-m_1^2}}{E\,u_1^3},
\qquad
d_2(E)=\frac{\sqrt{E^2/4-m_2^2}}{E\,u_2^3}.
\label{eq:d1d2_iso_drop}
\end{align}
Hence the imaginary part of one-loop matrix element keeps the same matrix form,
\begin{align}
\Im M_{\rm loop}^{(1)}(E)
=
\frac{1}{16\pi}\,
M_{\rm tree}\,
\diag\!\big(d_1(E),d_{12}(E),d_2(E)\big)\,
M_{\rm tree},
\label{eq:ImM_iso_drop}
\end{align}
but now all anisotropic angular integrals collapse to explicit functions of the two isotropic velocities \(u_1\) and \(u_2\). In particular, the angular-space operator becomes diagonal in the ordinary partial-wave basis, and at tree level only the \(s\)-wave sector is populated.

\subsection{Coupled-channel eigenvalues}

It is now useful to pass from the channel matrix to the quantity on which elastic unitarity acts directly, namely the phase-space-rescaled amplitude in the $Y_{00}$ sector. Using the diagonal phase-space matrix $D(E)$ introduced above, we obtain
\begin{align}
\widetilde a_{\rm tree}(E)
=
D(E)^{1/2}\,a^{\rm tree}_{00}(E)\,D(E)^{1/2}
=
-\frac{1}{16\pi}
\begin{pmatrix}
\lambda_1 d_1(E) & 0 & \lambda_3\sqrt{d_1(E)d_2(E)}\\
0 & \lambda_3 d_{12}(E) & 0\\
\lambda_3\sqrt{d_1(E)d_2(E)} & 0 & \lambda_2 d_2(E)
\end{pmatrix}.
\label{eq:atilde_tree_main}
\end{align}
Written in this form, the structure of the problem becomes immediate. The mixed channel $\beta=(\phi_1,\phi_2)$ is already isolated from the coupled identical-particle sector, so it contributes a single eigenvalue on its own,
\begin{align}
\Lambda_\beta(E)
=
-\frac{\lambda_3 d_{12}(E)}{16\pi}.
\label{eq:Lambda_beta_main}
\end{align}
The remaining two eigenvalues arise from diagonalizing the coupled $(\alpha,\gamma)$ block, and are therefore
\begin{align}
\Lambda_\pm(E)
=
-\frac{1}{32\pi}
\left[
\lambda_1 d_1(E)+\lambda_2 d_2(E)
\pm
\sqrt{
\big(\lambda_1 d_1(E)-\lambda_2 d_2(E)\big)^2
+
4\lambda_3^2 d_1(E)d_2(E)
}
\right].
\label{eq:Lambda_pm_main}
\end{align}
The perturbative elastic unitarity conditions therefore take the compact form
\begin{align}
|\Lambda_\beta(E)|\le \frac12,
\qquad
|\Lambda_\pm(E)|\le \frac12.
\label{eq:tree_unitarity_bounds_main}
\end{align}
The quantities bounded by unitarity are not the individual entries of the original channel matrix, but the eigenvalues of the amplitude after it has been dressed by the appropriate phase-space factors. 

\paragraph{Isotropic unequal-velocity case of the eigenvalue bounds.}

For
\begin{align}
\mC_1=u_1^2 I,
\qquad
\mC_2=u_2^2 I,
\qquad
u_1\neq u_2,
\end{align}
the angular dependence disappears and the exact elastic problem reduces to the ordinary coupled $s$-wave system. The phase-space factors are
\begin{align}
&d_1(E)
=
\frac{\sqrt{E^2/4-m_1^2}}{E\,u_1^3}
=
\frac{1}{2u_1^3}\sqrt{1-\frac{4m_1^2}{E^2}},
\nonumber\\
&d_2(E)
=
\frac{\sqrt{E^2/4-m_2^2}}{E\,u_2^3}
=
\frac{1}{2u_2^3}\sqrt{1-\frac{4m_2^2}{E^2}},
\label{eq:d12_iso_eig_drop}
\end{align}
and
\begin{align}
d_{12}(E)
=
\frac{2p_{12}(E)}
{u_1^2E_2(E)+u_2^2E_1(E)},
\qquad
E_i(E)=\sqrt{m_i^2+u_i^2p_{12}(E)^2},
\label{eq:d12mix_iso_eig_drop}
\end{align}
with $p_{12}(E)$ determined by the equation
\begin{align}
E=\sqrt{m_1^2+u_1^2p^2(E)}+\sqrt{m_2^2+u_2^2p^2(E)}.
\end{align}
The rescaled tree-level eigenvalues are therefore
\begin{align}
\Lambda_\beta(E)
=
-\frac{\lambda_3\,d_{12}(E)}{16\pi},
\label{eq:Lambda_beta_iso_drop}
\end{align}
and
\begin{align}
&\Lambda_\pm(E)\nonumber\\
&=
-\frac{1}{32\pi}
\left[
\lambda_1 d_1(E)+\lambda_2 d_2(E)
\pm
\sqrt{\big(\lambda_1 d_1(E)-\lambda_2 d_2(E)\big)^2
+4\lambda_3^2d_1(E)d_2(E)}\;
\right].
\label{eq:Lambda_pm_iso_drop}
\end{align}
Hence the exact elastic bounds are
\begin{align}
|\Lambda_\beta(E)|\le \frac12,
\qquad
|\Lambda_\pm(E)|\le \frac12.
\label{eq:unitarity_iso_drop}
\end{align}

In the physically relevant regime $\lambda_1,\lambda_2>0$, one has
$|\Lambda_-(E)|\le |\Lambda_+(E)|$, so the coupled identical-particle sector is controlled by $\Lambda_+(E)$ together with the mixed-channel eigenvalue $\Lambda_\beta(E)$. It is useful to note that
\begin{align}
\frac{d}{dE}d_i(E)
=
\frac{2m_i^2}{u_i^3E^3\sqrt{1-4m_i^2/E^2}}
\ge 0,
\label{eq:di_monotone_iso_drop}
\end{align}
while, using $p$ as the independent variable,
\begin{align}
\frac{d}{dp}d_{12}(p)
=
\frac{2}{\big(u_1^2E_2+u_2^2E_1\big)^2}
\left(
\frac{u_1^2m_2^2}{E_2}
+
\frac{u_2^2m_1^2}{E_1}
\right)
\ge 0.
\label{eq:d12_monotone_iso_drop}
\end{align}
Thus all three weights grow monotonically from threshold to their ultraviolet plateaux,
\begin{align}
d_1^\infty=\frac{1}{2u_1^3},
\qquad
d_2^\infty=\frac{1}{2u_2^3},
\qquad
d_{12}^\infty=\frac{2}{u_1u_2(u_1+u_2)}.
\label{eq:plateaux_iso_drop}
\end{align}
The strongest all-energy elastic bounds are therefore the ultraviolet ones.

For the coupled $(\phi_1\phi_1,\phi_2\phi_2)$ sector this gives the sharp condition
\begin{align}
\frac{\lambda_1}{u_1^3}
+
\frac{\lambda_2}{u_2^3}
+
\sqrt{
\left(\frac{\lambda_1}{u_1^3}-\frac{\lambda_2}{u_2^3}\right)^2
+
\frac{4\lambda_3^2}{u_1^3u_2^3}
}
\le 32\pi.
\label{eq:sharp_uv_pm_iso_drop}
\end{align}
Equivalently, this single inequality may be written as the manifestly sharp set
\begin{align}
\lambda_1\le 16\pi u_1^3,
\qquad
\lambda_2\le 16\pi u_2^3,
\qquad
\lambda_3^2\le
\big(16\pi u_1^3-\lambda_1\big)\big(16\pi u_2^3-\lambda_2\big).
\label{eq:sharp_uv_equiv_iso_drop}
\end{align}
For the mixed channel one finds independently
\begin{align}
|\lambda_3|\le 4\pi\,u_1u_2(u_1+u_2).
\label{eq:sharp_uv_beta_iso_drop}
\end{align}
Hence the portal coupling is constrained by the sharp combined bound
\begin{align}
|\lambda_3|
\le
\min\!\Bigg\{
4\pi\,u_1u_2(u_1+u_2),
\,
\sqrt{\big(16\pi u_1^3-\lambda_1\big)\big(16\pi u_2^3-\lambda_2\big)}
\Bigg\}.
\label{eq:portal_sharp_combined_iso_drop}
\end{align}

These inequalities may be read directly as lower bounds on the microscopic speeds. In particular,
\begin{align}
u_1\ge \left(\frac{\lambda_1}{16\pi}\right)^{1/3},
\qquad
u_2\ge \left(\frac{\lambda_2}{16\pi}\right)^{1/3},
\label{eq:speed_lb_diag_iso_drop}
\end{align}
and, for fixed $u_1$ with $16\pi u_1^3>\lambda_1$,
\begin{align}
u_2
&\ge
\left[
\frac{1}{16\pi}
\left(
\lambda_2+\frac{\lambda_3^2}{16\pi u_1^3-\lambda_1}
\right)
\right]^{1/3},
\label{eq:speed_lb_coupled_iso_drop}
\\[1mm]
u_2
&\ge
\frac{-u_1+\sqrt{u_1^2+\dfrac{|\lambda_3|}{\pi u_1}}}{2},
\label{eq:speed_lb_mixed_iso_drop}
\end{align}
with the second inequality coming from the mixed channel. The message is simple and sharp: once one speed approaches its individual limit, the allowed portal interaction collapses, unless the other speed is increased accordingly.

If one of the ultraviolet inequalities in
\eqref{eq:sharp_uv_equiv_iso_drop}--\eqref{eq:sharp_uv_beta_iso_drop}
is violated, perturbative unitarity fails at a finite energy. Because the weights are monotone, the corresponding breakdown scale is unique. For the individual identical channels one may solve it explicitly:
\begin{align}
E_{i,\max}
=
\frac{2m_i}{\sqrt{1-\left(\dfrac{16\pi u_i^3}{\lambda_i}\right)^2}},
\qquad
\lambda_i>16\pi u_i^3.
\label{eq:Eimax_iso_drop}
\end{align}
For the mixed channel and for the coupled $\Lambda_+$ sector, the maximal energies are defined implicitly by
\begin{align}
d_{12}(E_{\beta,\max})=\frac{8\pi}{|\lambda_3|},
\label{eq:Ebetamax_iso_drop}
\end{align}
and
\begin{gather}
\frac12
\Big[
\lambda_1 d_1(E_{+,\max})+\lambda_2 d_2(E_{+,\max})
+\\
\sqrt{\big(\lambda_1 d_1(E_{+,\max})-\lambda_2 d_2(E_{+,\max})\big)^2
+4\lambda_3^2d_1(E_{+,\max})d_2(E_{+,\max})}
\Big]
=
8\pi.
\label{eq:Epmax_iso_drop}
\end{gather}
In particular, the slow-velocity limit is strongly constrained: at fixed nonzero couplings one cannot send $u_i\to 0$ while remaining in the perturbative elastic regime. Rather, one must scale
\begin{align}
\lambda_i=O(u_i^3),
\qquad
\lambda_3=O\!\big(u_1u_2(u_1+u_2)\big),
\label{eq:smallu_scaling_iso_drop}
\end{align}
if a finite all-energy elastic window is to survive.

\section{Weak anisotropy}
\label{sec:weakk}

\subsection{Weak anisotropy and the first angular mixing}

It is instructive to examine the anisotropic problem in a neighborhood of the isotropic point. In that regime the kinetic matrices may be written as small deformations of isotropic propagation cones,
\begin{align}
\mC_i=u_i^2\big(I+\varepsilon_i\Delta_i\big),
\qquad
|\varepsilon_i|\ll 1,
\qquad
\Delta_i^T=\Delta_i.
\label{eq:C_weak_main}
\end{align}
Here $u_i$ is the isotropic reference speed of the $i$th field, while $\varepsilon_i\Delta_i$ measures the departure from spherical symmetry. The directional quadratic form then becomes
\begin{align}
\omega_i(\hat{\mathbf n})
=
u_i^2\big(1+\varepsilon_i q_i(\hat{\mathbf n})\big),
\qquad
q_i(\hat{\mathbf n})\equiv \hat{\mathbf n}^T\Delta_i\hat{\mathbf n}.
\label{eq:q_i_main}
\end{align}
Thus the entire weak-anisotropy problem is encoded, at first order, in the angular structure of the scalar function $q_i(\hat{\mathbf n})$ on the sphere.

The next step is to separate the isotropic and genuinely anisotropic pieces of $\Delta_i$. Writing
\begin{align}
\Delta_i=\frac{\tr \Delta_i}{3}\,I+Q_i,
\qquad
\tr Q_i=0,
\label{eq:Delta_decomp_main}
\end{align}
one finds that the angular dependence of $q_i$ contains only a monopole and a quadrupole,
\begin{align}
q_i(\hat{\mathbf n})
=
\frac{\tr\Delta_i}{3}
+
\sum_{m=-2}^{2}q_{i,2m}\,Y_{2m}(\hat{\mathbf n}).
\label{eq:q_i_harmonics_main}
\end{align}
This decomposition already reveals the basic structure of the perturbation. The trace part merely rescales the angular average of the phase-space measure, whereas the traceless part has purely quadrupolar angular structure and is therefore the first source of genuine angular mixing.

For this reason the first-order correction to any phase-space kernel  takes the universal form
\begin{align}
\delta g(E,\hat{\mathbf n})
=
\bar g_{00}(E)\,Y_{00}(\hat{\mathbf n})
+
\sum_{m=-2}^{2}\bar g_{2m}(E)\,Y_{2m}(\hat{\mathbf n}).
\label{eq:delta_g_main}
\end{align}
 The scalar component shifts the averaged $s$-wave normalization, while the quadrupole component is the first term that can couple different partial waves. Indeed, inserting the $J=2$ part into the angular operator \eqref{eq:G_operator_main} yields the selection rule
\begin{align}
\delta G_{l'm';lm}(E)\neq 0
\qquad\Longrightarrow\qquad
l'=l,\ l\pm 2,
\label{eq:selection_rule_weak_main}
\end{align}
and, in particular,
\begin{align}
\delta G_{2m;00}(E)=\frac{\bar g_{2m}(E)}{\sqrt{4\pi}}.
\label{eq:s_d_mixing_main}
\end{align}
Thus, near isotropy, the first nontrivial angular effect is sharply constrained: the $s$-wave mixes first with the $d$-wave, and with no other sector.

For an identical channel this structure becomes especially transparent, because the phase-space kernel admits a simple closed expansion. Substituting $\mC_i=u_i^2(I+\varepsilon_i\Delta_i)$ into the exact identical-particle expression gives
\begin{align}
g_i(E,\hat{\mathbf n})
=
\frac{k_i(E)}{E\,u_i^3}
\big(1+\varepsilon_i q_i(\hat{\mathbf n})\big)^{-3/2},
\qquad
k_i(E)\equiv \sqrt{\frac{E^2}{4}-m_i^2},
\label{eq:gi_weak_app}
\end{align}
and therefore
\begin{align}
g_i(E,\hat{\mathbf n})
=
g_i^{(0)}(E)
\left[
1
-\frac{3\varepsilon_i}{2}q_i(\hat{\mathbf n})
+\frac{15\varepsilon_i^2}{8}q_i(\hat{\mathbf n})^2
+O(\varepsilon_i^3)
\right],
\qquad
g_i^{(0)}(E)=\frac{k_i(E)}{E\,u_i^3}.
\label{eq:gi_expand_app}
\end{align}
This formula makes the geometry of the perturbation easy to read: the weak anisotropy enters only through the angular profile $q_i(\hat{\mathbf n})$, while the isotropic factor $g_i^{(0)}(E)$ carries the entire energy dependence of the undeformed channel.

The angular average can be obtained either directly from \eqref{eq:gi_expand_app}, or finding it from the exact determinant formula. For the latter, one first finds
\begin{align}
d_i(E)
=
d_i^{(0)}(E)
\left[
1
-\frac{\varepsilon_i}{2}\tr\Delta_i
+
\varepsilon_i^2
\left(
\frac18(\tr\Delta_i)^2+\frac14\tr(\Delta_i^2)
\right)
+O(\varepsilon_i^3)
\right].
\label{eq:di_weak_main}
\end{align}
This expression is useful because it separates the isotropic shift from the genuinely anisotropic one in a basis-independent way. In particular, if the deformation is traceless, then the averaged identical-channel factor receives no correction at linear order:
\begin{align}
\tr\Delta_i=0
\qquad\Longrightarrow\qquad
d_i(E)
=
d_i^{(0)}(E)
\left[
1+\frac{\varepsilon_i^2}{4}\tr(\Delta_i^2)+O(\varepsilon_i^3)
\right].
\label{eq:traceless_weak_main}
\end{align}
The leading observable effect is then not a shift of the averaged $s$-wave weight, but precisely the off-diagonal $s$--$d$ mixing encoded in \eqref{eq:s_d_mixing_main}. In other words, for a purely traceless perturbation the first signal of anisotropy appears in angular mixing before it appears in the averaged phase space.

The mixed channel is slightly richer, because it is sensitive not only to the deformation tensors themselves but also to the mismatch of the underlying isotropic speeds. To make this explicit, write
\begin{align}
\omega_1(\hat{\mathbf n})=u_1^2(1+\eta_1),
\qquad
\omega_2(\hat{\mathbf n})=u_2^2(1+\eta_2),
\qquad
\eta_i=\varepsilon_i q_i(\hat{\mathbf n}),
\label{eq:omega12_weak_app}
\end{align}
and let $p_0(E)$ denote the isotropic center-of-mass momentum determined by
\begin{align}
E=\bar E_1+\bar E_2,
\qquad
\bar E_1=\sqrt{m_1^2+u_1^2p_0^2},
\qquad
\bar E_2=\sqrt{m_2^2+u_2^2p_0^2}.
\label{eq:p0_def_app}
\end{align}
The weak anisotropy shifts the on-shell momentum away from $p_0$. Writing
\begin{align}
p=p_0+\delta p,
\end{align}
and imposing the on-shell condition to first order, $\delta E_1+\delta E_2=0$, gives
\begin{align}
\delta p(\hat{\mathbf n})
=
-\frac{p_0}{2}
\frac{
\bar E_2u_1^2\,\eta_1(\hat{\mathbf n})
+
\bar E_1u_2^2\,\eta_2(\hat{\mathbf n})
}{
D_0
},
\qquad
D_0\equiv \bar E_2u_1^2+\bar E_1u_2^2.
\label{eq:deltap_app}
\end{align}
Substituting this correction into the exact mixed kernel,
\begin{align}
g_{12}(E,\hat{\mathbf n})
=
\frac{2p(E,\hat{\mathbf n})}{
E_2(E,\hat{\mathbf n})\omega_1(\hat{\mathbf n})
+
E_1(E,\hat{\mathbf n})\omega_2(\hat{\mathbf n})},
\label{eq:g12_def_app}
\end{align}
one obtains the linearized form
\begin{align}
g_{12}(E,\hat{\mathbf n})
=
g_{12}^{(0)}(E)
\left[
1+C_1(E)\eta_1(\hat{\mathbf n})+C_2(E)\eta_2(\hat{\mathbf n})
\right]
+O(\varepsilon^2),
\label{eq:g12_mixed_main}
\end{align}
where
\begin{align}
g_{12}^{(0)}(E)=\frac{2p_0}{D_0},
\label{eq:g12zero_app}
\end{align}
and
\begin{align}
C_1(E)
&=
-\frac{3\bar E_2u_1^2}{2D_0}
-
\frac{(u_2^2-u_1^2)p_0^2u_1^2u_2^2}{2D_0^2},
\nonumber\\
C_2(E)
&=
-\frac{3\bar E_1u_2^2}{2D_0}
+
\frac{(u_2^2-u_1^2)p_0^2u_1^2u_2^2}{2D_0^2}.
\label{eq:C1C2_app}
\end{align}

These coefficients admit a clean interpretation. The first term in each of $C_1(E)$ and $C_2(E)$ is the direct response of the phase-space denominator to a deformation of the corresponding propagation cone. The second term is indirect: it comes from the fact that the deformed kinematics also shifts the on-shell momentum itself. This second contribution disappears when $u_1=u_2$, as it must, because in that case the isotropic reference problem no longer distinguishes the two sectors kinematically.

In the weakly anisotropic regime, the universal angular content of the deformation is monopolar plus quadrupolar, and the first genuinely new partial-wave effect is $s$--$d$ mixing. For identical channels a traceless deformation leaves the averaged $s$-wave measure unchanged at linear order, so the angular mixing is the leading signal. For the mixed channel the same quadrupolar structure persists, but its coefficient functions also retain memory of the unequal isotropic speeds through the energy-dependent responses $C_1(E)$ and $C_2(E)$. This is the simplest setting in which one can see, in fully explicit form, how relative anisotropy and unequal propagation speeds cooperate in the elastic problem.

\section{One-loop effective potential and renormalization}
\label{sec:effpot}

We now pass from scattering kinematics to the background-field problem for the same two-scalar theory. The objective is to understand how the mismatch of spatial kinetic tensors reappears in the one-loop effective potential. At this stage the central issue is no longer the geometry of two-body phase space, but the structure of the quadratic fluctuation operator in a homogeneous classical background. We therefore begin from the microscopic Lagrangian \eqref{eq:model_quartic_main}.
To construct the effective potential, we split each field into a constant background and a quantum fluctuation,
\begin{align}
\phi_1=\varphi_1+\eta_1,
\qquad
\phi_2=\varphi_2+\eta_2,
\label{eq:bg_split_main}
\end{align}
where $\varphi_i$ are homogeneous classical expectation values and $\eta_i$ denote the fluctuating fields to be integrated out. The tree-level potential evaluated on the background is then
\begin{align}
V_{\rm tree}(\varphi_1,\varphi_2)
=
\frac12 m_1^2\varphi_1^2+\frac12 m_2^2\varphi_2^2
+\frac{\lambda_1}{24}\varphi_1^4
+\frac{\lambda_2}{24}\varphi_2^4
+\frac{\lambda_3}{4}\varphi_1^2\varphi_2^2.
\label{eq:Vtree_main}
\end{align}
Already at this classical level the potential has the familiar quartic two-field form. What is new in the present setting is that the fluctuations propagate with different spatial quadratic forms, so the one-loop determinant will retain detailed memory of the matrices $\mC_1$ and $\mC_2$.

The background enters the quadratic fluctuation operator through the effective mass matrix. Its diagonal entries are some new masses combinations of the two fields in the chosen background, while the off-diagonal entry is induced by the portal interaction. Explicitly,
\begin{align}
M_1^2
=
m_1^2+\frac{\lambda_1}{2}\varphi_1^2+\frac{\lambda_3}{2}\varphi_2^2,
\qquad
M_2^2
=
m_2^2+\frac{\lambda_2}{2}\varphi_2^2+\frac{\lambda_3}{2}\varphi_1^2,
\qquad
s=\lambda_3\varphi_1\varphi_2.
\label{eq:M1M2s_main}
\end{align}
The quantities $M_1^2$ and $M_2^2$ are the background-dependent diagonal masses, whereas $s$ measures the mixing between the two fluctuation sectors. These three combinations will organize the full one-loop problem: they determine the eigenvalues of the quadratic operator, control the structure of the determinant, and ultimately separate the diagonal and genuinely mixed contributions to the effective potential.

After Wick rotation, the quadratic Euclidean operator is
\begin{align}
\mathcal D_E(k)
=
\begin{pmatrix}
k_4^2+Q_1(\mathbf k)+M_1^2 & s\\
s & k_4^2+Q_2(\mathbf k)+M_2^2
\end{pmatrix},
\qquad
Q_i(\mathbf k)\equiv \mathbf k^T\mC_i\mathbf k.
\label{eq:DE_main}
\end{align}
The one-loop effective potential is
\begin{align}
V_1(\varphi_1,\varphi_2)
=
\frac12\int\frac{d^4k_E}{(2\pi)^4}\,\ln\det\mathcal D_E(k).
\label{eq:V1_det_main}
\end{align}
Diagonalizing the $2\times2$ operator in field space gives
\begin{align}
V_1(\varphi_1,\varphi_2)
=
\frac12\int\frac{d^3\mathbf k}{(2\pi)^3}\,
\big[\Omega_+(\mathbf k)+\Omega_-(\mathbf k)\big],
\label{eq:V1_exact_main}
\end{align}
where
\begin{align}
\Omega_\pm^2(\mathbf k)
=
\frac{
Q_1(\mathbf k)+Q_2(\mathbf k)+M_1^2+M_2^2
\pm
\sqrt{\big(Q_1(\mathbf k)-Q_2(\mathbf k)+M_1^2-M_2^2\big)^2+4s^2}
}{2}.
\label{eq:Omega_pm_main}
\end{align}
This representation is exact, but it also makes the obstruction to a simple closed form manifest: unless $\mC_1=\mC_2$, the mixing angle depends on momentum, so no momentum-independent field redefinition diagonalizes the determinant mode by mode.

A particularly useful benchmark is the equal-anisotropy locus,
\begin{align}
\mC_1=\mC_2\equiv \mC,
\label{eq:equalC_main}
\end{align}
for which $\Omega_\pm^2(\mathbf k)=\mathbf k^T\mC\,\mathbf k+w_\pm$, where
\begin{align}
    w_{ \pm}=\frac{M_1^2+M_2^2 \pm \sqrt{\left(M_1^2-M_2^2\right)^2+4 s^2}}{2},
\end{align}
and the determinant reduces to the ordinary Coleman--Weinberg form after a spatial rescaling:
\begin{align}
V_1\big|_{\mC_1=\mC_2=\mC}
=
\sum_{\sigma=\pm}
\frac{w_\sigma^2}{64\pi^2\sqrt{\det\mC}}
\left[
\ln\!\left(\frac{w_\sigma}{\mu^2\bar u_{\mC}^2}\right)-\frac32
\right],
\qquad
\bar u_{\mC}^2\equiv (\det\mC)^{1/3}.
\label{eq:V1_equalC_main}
\end{align}
Away from this locus, the exact potential retains the nonlocal structure of \eqref{eq:V1_exact_main}.

\subsection{Local quartic truncation: determinant expansion and loop integrals}

For renormalization one does not need the full nonlocal determinant. The ultraviolet counterterms and one-loop beta functions are already fixed once one isolates the local part through quartic order in the backgrounds. The cleanest way to do this is to split the quadratic fluctuation operator into a \textit{reference operator} $\mathcal D_0$, evaluated at vanishing background, and a \textit{background insertion matrix} $\mathcal V$ that collects the quadratic and mixed background couplings. Concretely, write
\begin{align}
\mathcal D_E(k)=\mathcal D_0(k)+\mathcal V,
\qquad
\mathcal D_0(k)
=
\begin{pmatrix}
D_1 & 0\\
0 & D_2
\end{pmatrix},
\qquad
\mathcal V
=
\begin{pmatrix}
U_1 & s\\
s & U_2
\end{pmatrix},
\label{eq:D0V_app}
\end{align}
with
\begin{align}
D_1=k_4^2+Q_1(\mathbf k)+m_1^2,
\qquad
D_2=k_4^2+Q_2(\mathbf k)+m_2^2,
\label{eq:D1D2_app}
\end{align}
and
\begin{align}
U_1=\frac{\lambda_1}{2}\varphi_1^2+\frac{\lambda_3}{2}\varphi_2^2,
\qquad
U_2=\frac{\lambda_3}{2}\varphi_1^2+\frac{\lambda_2}{2}\varphi_2^2.
\label{eq:U1U2_main}
\end{align}
Then
\begin{align}
\ln\det(\mathcal D_0+\mathcal V)
=
\ln\det\mathcal D_0+\Tr\ln(1+\mathcal D_0^{-1}\mathcal V),
\label{eq:logdet_app}
\end{align}
and since $\mathcal V=O(\varphi^2)$,
\begin{align}
V_1
=
\frac12\int\frac{d^dk}{(2\pi)^d}\Tr(\mathcal D_0^{-1}\mathcal V)
-\frac14\int\frac{d^dk}{(2\pi)^d}
\Tr(\mathcal D_0^{-1}\mathcal V\,\mathcal D_0^{-1}\mathcal V)
+O(\varphi^6),
\label{eq:V1_expand_app}
\end{align}
with
\begin{align}
\Tr(\mathcal D_0^{-1}\mathcal V)
=
\frac{U_1}{D_1}+\frac{U_2}{D_2},
\qquad
\Tr(\mathcal D_0^{-1}\mathcal V\,\mathcal D_0^{-1}\mathcal V)
=
\frac{U_1^2}{D_1^2}+\frac{U_2^2}{D_2^2}+\frac{2s^2}{D_1D_2}.
\label{eq:traces_app}
\end{align}
Because $U_1$, $U_2$, and $s$ are already quadratic in the background fields, the terms displayed above are precisely the ones needed to reconstruct the local effective potential through quartic order. In the ultraviolet/local expansion around nonzero masses, no higher power of $\mathcal V$ contributes to the renormalizable local sector.

In dimensional regularization with $d=4-2\varepsilon$ and
\begin{align}
\frac{1}{\bar\varepsilon}
=
\frac{1}{\varepsilon}-\gamma_E+\ln(4\pi),
\label{eq:barepsilon_app}
\end{align}
the single-propagator integrals reduce after spatial rescaling to
\begin{align}
\int\frac{d^dk}{(2\pi)^d}\frac{1}{D_i}
&=
-\frac{m_i^2}{16\pi^2\Pi_i}
\left[
\frac{1}{\bar\varepsilon}
-\ln\!\left(\frac{m_i^2}{\mu^2\bar u_i^2}\right)+1
\right],
\nonumber\\
\int\frac{d^dk}{(2\pi)^d}\frac{1}{D_i^2}
&=
\frac{1}{16\pi^2\Pi_i}
\left[
\frac{1}{\bar\varepsilon}
-\ln\!\left(\frac{m_i^2}{\mu^2\bar u_i^2}\right)
\right],
\label{eq:singleprop_app}
\end{align}
where
\begin{align}
\Pi_i\equiv \sqrt{\det\mC_i},
\qquad
\bar u_i^2\equiv (\det\mC_i)^{1/3}.
\label{eq:Pi_uibar_main}
\end{align}
Let us note once again that throughout this subsection we use a dimensional-regularization prescription in which the anisotropic spatial tensors are kept as physical $3\times 3$ objects. Concretely, for each diagonal denominator $D_i$ and for the Feynman-parameterized mixed denominator, we first perform the corresponding real three-dimensional linear change of variables that trivializes the spatial quadratic form. The Jacobian of this transformation produces the factors $\Pi_i^{-1}=1/\sqrt{\det\mC_i}$ or $\Xi(x)^{-1}=1/\sqrt{\det\mC(x)}$. Only the remaining rotationally invariant Euclidean loop integral is then analytically continued to $d=4-2\varepsilon$. In this sense, $\bar u_i^2=(\det\mC_i)^{1/3}$ and $\bar u(x)^2=(\det\mC(x))^{1/3}$ are physical three-dimensional geometric scales inherited from the spatial rescaling, rather than quantities continued away from three spatial dimensions.

Next, the genuinely mixed bubble is
\begin{align}
I_{12}\equiv \int\frac{d^dk}{(2\pi)^d}\frac{1}{D_1D_2}.
\label{eq:I12_app}
\end{align}
Using a Feynman parameter,
\begin{align}
\frac{1}{D_1D_2}
=
\int_0^1 dx\,\frac{1}{[xD_1+(1-x)D_2]^2},
\label{eq:feynman_I12_app}
\end{align}
with
\begin{align}
&\mC(x)\equiv x\mC_1+(1-x)\mC_2,\nonumber
\\
&\Xi(x)\equiv \sqrt{\det \mC(x)},
\nonumber\\
&M(x)^2\equiv xm_1^2+(1-x)m_2^2,
\nonumber\\
&\bar u(x)^2\equiv (\det \mC(x))^{1/3},
\label{eq:Cx_Mx_main}
\end{align}
gives
\begin{align}
I_{12}
=
\frac{1}{16\pi^2}
\int_0^1 dx\,\frac{1}{\Xi(x)}
\left[
\frac{1}{\bar\varepsilon}
-\ln\!\left(\frac{M(x)^2}{\mu^2\bar u(x)^2}\right)
\right]
=
\frac{1}{16\pi^2}
\left[
\frac{\mathcal J_{12}}{\bar\varepsilon}
-
K_{12}^{\rm anisotropic}
\right],
\label{eq:I12_final_app}
\end{align}
where
\begin{align}
K_{12}^{\rm anisotropic}(m_1,m_2;\mC_1,\mC_2;\mu)
=
\int_0^1 dx\,\frac{1}{\Xi(x)}
\ln\!\left(\frac{M(x)^2}{\mu^2\bar u(x)^2}\right),
\label{eq:K12_main}
\end{align}
and
\begin{align}
\mathcal J_{12}\equiv \int_0^1 dx\,\frac{1}{\Xi(x)}.
\label{eq:J12_main}
\end{align}
Already at the level of local ultraviolet data, the anisotropy therefore splits into two different structures: self-contractions carry $\Pi_i^{-1}$, while mixed contractions carry $\mathcal J_{12}$.

Substituting these integrals into the determinant expansion yields the renormalizable local potential,
\begin{align}
V_{\rm eff}^{\rm local}(\varphi_1,\varphi_2)
&=
V_{\rm tree}(\varphi_1,\varphi_2)
+\sum_{i=1}^{2}
\frac{1}{64\pi^2\Pi_i}
\left[
2m_i^2U_i\left(\ln\!\frac{m_i^2}{\mu^2\bar u_i^2}-1\right)
+U_i^2\ln\!\frac{m_i^2}{\mu^2\bar u_i^2}
\right]
\nonumber\\
&\qquad
+\frac{s^2}{32\pi^2}\,
K_{12}^{\rm anisotropic}(m_1,m_2;\mC_1,\mC_2;\mu)
+\cO(\varphi^6).
\label{eq:Veff_local_main}
\end{align}

\subsection{Counterterms and one-loop beta functions}

The ultraviolet structure can now be read off directly from the local expansion. Once the divergent parts of the diagonal tadpoles, diagonal bubbles, and the genuinely mixed bubble are assembled, the required counterterms take the form
\begin{align}
\delta\lambda_1
&=
\frac{3}{32\pi^2}\frac{1}{\bar\varepsilon}
\left(
\frac{\lambda_1^2}{\Pi_1}+\frac{\lambda_3^2}{\Pi_2}
\right),
\nonumber\\
\delta\lambda_2
&=
\frac{3}{32\pi^2}\frac{1}{\bar\varepsilon}
\left(
\frac{\lambda_2^2}{\Pi_2}+\frac{\lambda_3^2}{\Pi_1}
\right),
\nonumber\\
\delta\lambda_3
&=
\frac{1}{\bar\varepsilon}
\left[
\frac{\lambda_1\lambda_3}{32\pi^2\Pi_1}
+\frac{\lambda_2\lambda_3}{32\pi^2\Pi_2}
+\frac{\lambda_3^2}{8\pi^2}\,\mathcal J_{12}
\right],
\label{eq:deltalambdas_app}
\end{align}
while the mass renormalization is governed by
\begin{align}
\delta m_1^2
&=
\frac{1}{32\pi^2}\frac{1}{\bar\varepsilon}
\left(
\frac{\lambda_1m_1^2}{\Pi_1}
+\frac{\lambda_3m_2^2}{\Pi_2}
\right),
\nonumber\\
\delta m_2^2
&=
\frac{1}{32\pi^2}\frac{1}{\bar\varepsilon}
\left(
\frac{\lambda_2m_2^2}{\Pi_2}
+\frac{\lambda_3m_1^2}{\Pi_1}
\right).
\label{eq:deltamasses_app}
\end{align}
These expressions make the pattern of the renormalization especially transparent. The self-contractions of each field are weighted by the corresponding geometric factors $\Pi_i^{-1}$, whereas the portal coupling also receives the genuinely mixed contribution proportional to $\mathcal J_{12}$. In this way the counterterm structure mirrors the same separation already encountered in the local effective potential: diagonal ultraviolet data are controlled by the individual ellipsoidal volumes, while the mixed sector retains explicit memory of the interpolating anisotropic geometry. After standard algebra one obtains the beta function corresponding to these couplings and masses
\begin{align}
\beta_{\lambda_1}
&=
\frac{3}{16\pi^2}
\left(
\frac{\lambda_1^2}{\Pi_1}+\frac{\lambda_3^2}{\Pi_2}
\right),
&
\beta_{m_1^2}
&=
\frac{1}{16\pi^2}
\left(
\frac{\lambda_1m_1^2}{\Pi_1}+\frac{\lambda_3m_2^2}{\Pi_2}
\right),
\nonumber\\[1mm]
\beta_{\lambda_2}
&=
\frac{3}{16\pi^2}
\left(
\frac{\lambda_2^2}{\Pi_2}+\frac{\lambda_3^2}{\Pi_1}
\right),
&
\beta_{m_2^2}
&=
\frac{1}{16\pi^2}
\left(
\frac{\lambda_2m_2^2}{\Pi_2}+\frac{\lambda_3m_1^2}{\Pi_1}
\right),
\nonumber\\[1mm]
\beta_{\lambda_3}
&=
\frac{\lambda_3}{16\pi^2}
\left(
\frac{\lambda_1}{\Pi_1}+\frac{\lambda_2}{\Pi_2}+4\lambda_3\,\mathcal J_{12}
\right).
\label{eq:betas_local_main}
\end{align}
At one loop there is no momentum-dependent self-energy in the unbroken quartic theory, hence no wavefunction renormalization and no running of the spatial tensors:
\begin{align}
\beta_{\mC_1}=0,
\qquad
\beta_{\mC_2}=0.
\label{eq:betaC_main}
\end{align}
For diagonal directional anisotropy one may also write
\begin{align}
\mathcal J_{12}
=
\int_0^1 dx\,
\frac{1}{
\prod_{a=1}^{3}\sqrt{x u_{1a}^2+(1-x)u_{2a}^2}
},
\label{eq:J12_diag_app}
\end{align}
and, introducing
\begin{align}
\Delta_a\equiv u_{1a}^2-u_{2a}^2,
\qquad
r_a\equiv \frac{u_{2a}^2}{\Delta_a},
\label{eq:Deltara_app}
\end{align}
one can express this in terms of Carlson's symmetric integral $R_F$:
\begin{align}
\mathcal J_{12}
=
\frac{2}{\sqrt{\Delta_1\Delta_2\Delta_3}}
\Big[
R_F(r_1,r_2,r_3)-R_F(r_1+1,r_2+1,r_3+1)
\Big],
\label{eq:Carlson_app}
\end{align}
where we assume that $\Delta_a \neq 0$ as well as that there a consistent real branch for $\sqrt{\Delta_1 \Delta_2 \Delta_3} $ exists.\footnote{Formula \eqref{eq:Carlson_app} is valid as a real expression when the three quantities $\Delta_a$ have the same sign. In mixed-sign cases the compact Carlson representation requires separate branch bookkeeping, and it is safer to use the integral form \eqref{eq:J12_diag_app}.}
Formula \eqref{eq:Carlson_app} makes clear that even the mixed ultraviolet weight admits a compact special-function representation once the spatial tensors are diagonalized.

\section{Nonpolynomial reorganization, scale invariance, and multiscale improvement}
\label{sec:msrg}

The exact determinant admits the following decomposition:
\begin{align}
V_1(\varphi_1,\varphi_2)
=
V_1^{(1)}(M_1^2)+V_1^{(2)}(M_2^2)+V_{\rm mix}(\varphi_1,\varphi_2),
\label{eq:V1split_main}
\end{align}
where the diagonal pieces are 
\begin{align}
    V_1^{(i)}\left(M_i^2\right)=\frac{1}{2} \int \frac{d^4 k_E}{(2 \pi)^4} \ln \mathcal{Q}_i(k),
\end{align}
and
\begin{align}
V_{\rm mix}
=
\frac12\int\frac{d^4k_E}{(2\pi)^4}
\ln\!\left(1-\frac{s^2}{\mathcal Q_1(k)\mathcal Q_2(k)}\right),
\qquad
\mathcal Q_i(k)\equiv k_4^2+Q_i(\mathbf k)+M_i^2.
\label{eq:Vmix_main}
\end{align}
In the convergence domain
\begin{align}
M_1^2>0,\qquad M_2^2>0,\qquad \frac{s^2}{M_1^2M_2^2}<1,
\label{eq:convergence_domain_main}
\end{align}
one may expand
\begin{align}
V_{\rm mix}
=
-\frac12\sum_{n=1}^{\infty}\frac{s^{2n}}{n}\,J_n(M_1^2,M_2^2;\mC_1,\mC_2).
\label{eq:Vmix_series_main}
\end{align}
The coefficients are
\begin{align}
J_n
\equiv
\int\frac{d^4k_E}{(2\pi)^4}
\frac{1}{\mathcal Q_1(k)^n\mathcal Q_2(k)^n}.
\label{eq:Jn_def_app}
\end{align}
For $n\ge 2$,
\begin{align}
J_n
=
\frac{1}{16\pi^2}
\frac{\Gamma(2n-2)}{\Gamma(n)^2}
\int_0^1 dx\,
\frac{x^{n-1}(1-x)^{n-1}}{\Xi(x)\,[\mathcal M(x)^2]^{2n-2}},
\qquad
\mathcal M(x)^2\equiv xM_1^2+(1-x)M_2^2,
\label{eq:Jn_formula_app}
\end{align}
whereas for $n=1$,
\begin{align}
J_1
=
\frac{1}{16\pi^2}
\int_0^1 dx\,\frac{1}{\Xi(x)}
\left[
\frac{1}{\bar\varepsilon}
-\ln\!\left(\frac{\mathcal M(x)^2}{\mu^2\bar u(x)^2}\right)
\right].
\label{eq:J1_formula_app}
\end{align}
Only $J_1$ is divergent; all $J_n$ with $n\ge 2$ are finite and begin at order $\varphi^8$ (what is true in the local quartic expansion counting around nonzero masses). This separation is one reason the mixed part of the determinant is so naturally adapted to a multiscale RG treatment.

$\,$

In the classically scale-invariant regime,
\begin{align}
m_1=m_2=0,
\label{eq:scale_inv_main}
\end{align}
the tree-level potential reduces to the purely quartic form
\begin{align}
V_0(\varphi_1,\varphi_2)
=
\frac{\lambda_1}{24}\varphi_1^4+\frac{\lambda_2}{24}\varphi_2^4+\frac{\lambda_3}{4}\varphi_1^2\varphi_2^2.
\label{eq:V0_SI_main}
\end{align}
It is then natural to parameterize field space in polar variables,
\begin{align}
\varphi_1=\rho\cos\theta,\qquad \varphi_2=\rho\sin\theta,
\label{eq:polar_main}
\end{align}
so that the search for a flat direction becomes a problem in the angular variable alone. In these variables one finds that the Gildener--Weinberg conditions are unchanged by anisotropy:
\begin{align}
\tan^2\theta_0=-\frac{3\lambda_3}{\lambda_2},
\qquad
\lambda_1\lambda_2=9\lambda_3^2,
\qquad
\lambda_3<0.
\label{eq:GW_main}
\end{align}
Thus the location of the flat ray is determined exactly as in the ordinary relativistic case. Along this ray, defined by $\varphi_i=\rho n_i$ with $n_1=\cos\theta_0$ and $n_2=\sin\theta_0$, the tree-level mass matrix develops one vanishing eigenvalue, corresponding to the classically flat direction, and one heavy mode with
\begin{align}
m_H^2(\rho)=-\lambda_3\rho^2.
\label{eq:mH_main}
\end{align}
Renormalization-group invariance then fixes the one-loop effective potential along the flat direction to take the universal Gildener--Weinberg form
\begin{align}
V_{\rm eff}(\rho)\big|_{\theta=\theta_0}
=
B\,\rho^4\left(\ln\!\frac{\rho^2}{v^2}-\frac12\right),
\label{eq:GW_final_main}
\end{align}
with coefficient \footnote{In the classically scale-invariant case one has $M_i^2(\rho)\propto \rho^2$ and $s(\rho)\propto \rho^2$, so the terms with $n\ge 2$ in \eqref{eq:Vmix_series_main} also contribute finite terms of order $\rho^4$. They do not, however, generate additional ultraviolet logarithms. Therefore the coefficient $B$ of the logarithmic term in \eqref{eq:GW_final_main} is fixed by the diagonal logarithms and by the mixed $J_1$ contribution alone; the higher $J_{n\ge 2}$ terms only modify the finite $\rho^4$ part of the potential.}
\begin{align}
B
=
\frac{1}{64\pi^2\rho^4}
\left[
\frac{M_1^4(\rho)}{\Pi_1}
+\frac{M_2^4(\rho)}{\Pi_2}
+2\,\mathcal J_{12}\,s(\rho)^2
\right]_{\theta=\theta_0,\ \mu=\mu_{\rm GW}}.
\label{eq:B_main}
\end{align}
The scalon mass therefore follows in the standard way:
\begin{align}
m_{\rm scalon}^2=8Bv^2.
\label{eq:scalon_mass_main}
\end{align}
The important point is that anisotropy does not move the flat direction itself; rather, it enters at one loop through the coefficient $B$, where the diagonal contributions are weighted by $\Pi_i^{-1}$ and the genuinely mixed contribution is weighted by $\mathcal J_{12}$.

For the isotropic-but-unequal case
\begin{align}
\mC_1=u_1^2 I,
\qquad
\mC_2=u_2^2 I,
\end{align}
the geometric factors entering \eqref{eq:B_main} reduce to
\begin{align}
\Pi_1=u_1^3,
\qquad
\Pi_2=u_2^3,
\qquad
\mathcal J_{12}=\frac{2}{u_1u_2(u_1+u_2)}.
\end{align}
Along the Gildener--Weinberg ray
one has
\begin{align}
M_1^2(\rho)
=
\frac{\rho^2}{2}\big(\lambda_1 n_1^2+\lambda_3 n_2^2\big),
\qquad
M_2^2(\rho)
=
\frac{\rho^2}{2}\big(\lambda_2 n_2^2+\lambda_3 n_1^2\big),
\qquad
s(\rho)=\lambda_3\rho^2 n_1n_2.
\end{align}
Therefore
\begin{align}
B_{\rm iso}
=
\frac{1}{64\pi^2}
\left[
\frac{\big(\lambda_1 n_1^2+\lambda_3 n_2^2\big)^2}{4u_1^3}
+
\frac{\big(\lambda_2 n_2^2+\lambda_3 n_1^2\big)^2}{4u_2^3}
+
\frac{4\lambda_3^2 n_1^2n_2^2}{u_1u_2(u_1+u_2)}
\right].
\label{eq:B_iso_drop}
\end{align}
Using the flat-direction relation $\lambda_1\lambda_2=9\lambda_3^2$, this can be rewritten as
\begin{align}
M_1^2(\rho)=\frac{\lambda_1}{3}n_1^2\rho^2,
\qquad
M_2^2(\rho)=\frac{\lambda_2}{3}n_2^2\rho^2,
\end{align}
and hence
\begin{align}
B_{\rm iso}
=
\frac{1}{64\pi^2}
\left[
\frac{\lambda_1^2 n_1^4}{9u_1^3}
+
\frac{\lambda_2^2 n_2^4}{9u_2^3}
+
\frac{4\lambda_3^2 n_1^2n_2^2}{u_1u_2(u_1+u_2)}
\right].
\end{align}

\subsection{Three-scale RG improvement}

The logarithmic structure of the one-loop potential suggests a multiscale subtraction scheme with three independent scales,
\begin{align}
\kappa_1,\qquad \kappa_2,\qquad \kappa_{12},
\label{eq:three_scales_main}
\end{align}
attached respectively to the field-1 diagonal determinant, the field-2 diagonal determinant, and the genuinely mixed logarithm. It is useful to think of these as three separate bookkeeping devices for three different ultraviolet sectors of the same theory. At one loop the renormalized potential satisfies
\begin{align}
\mathcal D_A V_{\rm eff}=0,
\qquad
A\in\{1,2,12\},
\label{eq:CS_partial_main}
\end{align}
where
\begin{align}
    \mathcal{D}_A=\kappa_A \frac{\partial}{\partial \kappa_A}+\sum_I \beta_I^{(A)} \frac{\partial}{\partial g_I},
\end{align}
with partial beta functions that sum to the ordinary ones in \eqref{eq:betas_local_main}. Concretely,
\begin{align}
\beta_{m_1^2}^{(1)}
=
\frac{\lambda_1m_1^2}{16\pi^2\Pi_1},
\qquad
\beta_{m_2^2}^{(1)}
=
\frac{\lambda_3m_1^2}{16\pi^2\Pi_1},
\label{eq:betamass1_app}
\end{align}
\begin{align}
\beta_{\lambda_1}^{(1)}
=
\frac{3\lambda_1^2}{16\pi^2\Pi_1},
\qquad
\beta_{\lambda_2}^{(1)}
=
\frac{3\lambda_3^2}{16\pi^2\Pi_1},
\qquad
\beta_{\lambda_3}^{(1)}
=
\frac{\lambda_1\lambda_3}{16\pi^2\Pi_1},
\label{eq:betalambda1_app}
\end{align}
and similarly
\begin{align}
\beta_{m_1^2}^{(2)}
=
\frac{\lambda_3m_2^2}{16\pi^2\Pi_2},
\qquad
\beta_{m_2^2}^{(2)}
=
\frac{\lambda_2m_2^2}{16\pi^2\Pi_2},
\label{eq:betamass2_app}
\end{align}
\begin{align}
\beta_{\lambda_1}^{(2)}
=
\frac{3\lambda_3^2}{16\pi^2\Pi_2},
\qquad
\beta_{\lambda_2}^{(2)}
=
\frac{3\lambda_2^2}{16\pi^2\Pi_2},
\qquad
\beta_{\lambda_3}^{(2)}
=
\frac{\lambda_2\lambda_3}{16\pi^2\Pi_2},
\label{eq:betalambda2_app}
\end{align}
while the mixed scale is even more selective:
\begin{align}
\beta_{\lambda_3}^{(12)}=\frac{\mathcal J_{12}}{4\pi^2}\lambda_3^2,
\qquad
\beta_{\lambda_1}^{(12)}=\beta_{\lambda_2}^{(12)}=\beta_{m_i^2}^{(12)}=0.
\label{eq:partial_beta12_main}
\end{align}
Thus the mixed logarithm renormalizes only the portal interaction. This separation is conceptually valuable, because it makes transparent which part of the flow comes from self-contractions and which part comes from the genuinely anisotropic mixed bubble.

The sequential one-loop solution is equally explicit. Define
\begin{align}
t_1=\ln\frac{\kappa_1}{\mu_0},
\qquad
t_2=\ln\frac{\kappa_2}{\mu_0},
\qquad
t_{12}=\ln\frac{\kappa_{12}}{\mu_0},
\label{eq:times_app}
\end{align}
and
\begin{align}
\mathfrak a_1\equiv \frac{1}{16\pi^2\Pi_1},
\qquad
\mathfrak a_2\equiv \frac{1}{16\pi^2\Pi_2},
\qquad
\mathfrak b\equiv \frac{\mathcal J_{12}}{4\pi^2}.
\label{eq:abcoeffs_app}
\end{align}
The first-stage flow is
\begin{align}
\lambda_1^{[1]}(t_1)
&=
\frac{\lambda_{1,0}}{1-3\mathfrak a_1\lambda_{1,0}t_1},
\nonumber\\
\lambda_3^{[1]}(t_1)
&=
\lambda_{3,0}\big(1-3\mathfrak a_1\lambda_{1,0}t_1\big)^{-1/3},
\nonumber\\
\lambda_2^{[1]}(t_1)
&=
\lambda_{2,0}
+\frac{3\lambda_{3,0}^2}{\lambda_{1,0}}
\left[
1-\big(1-3\mathfrak a_1\lambda_{1,0}t_1\big)^{1/3}
\right],
\label{eq:stage1lambda_app}
\end{align}
and
\begin{align}
m_1^{2[1]}(t_1)
&=
m_{1,0}^2\big(1-3\mathfrak a_1\lambda_{1,0}t_1\big)^{-1/3},
\nonumber\\
m_2^{2[1]}(t_1)
&=
m_{2,0}^2
+\frac{\lambda_{3,0}m_{1,0}^2}{\lambda_{1,0}}
\left[
1-\big(1-3\mathfrak a_1\lambda_{1,0}t_1\big)^{1/3}
\right].
\label{eq:stage1mass_app}
\end{align}
These formulas make the internal logic of the first flow easy to read: $\lambda_1$ runs with the familiar one-coupling rational form, $\lambda_3$ is dragged along with exponent $-1/3$, and $\lambda_2$ and $m_2^2$ receive induced portal contributions even before the second flow is turned on.

Using these as initial data for the second stage and defining
\begin{align}
A_2\equiv 1-3\mathfrak a_2\lambda_2^{[1]}t_2,
\label{eq:A2_app}
\end{align}
one finds
\begin{align}
\lambda_2^{[2]}=\frac{\lambda_2^{[1]}}{A_2},
\qquad
\lambda_3^{[2]}=\lambda_3^{[1]}A_2^{-1/3},
\qquad
\lambda_1^{[2]}
=
\lambda_1^{[1]}
+\frac{3(\lambda_3^{[1]})^2}{\lambda_2^{[1]}}
\left(1-A_2^{1/3}\right),
\label{eq:stage2lambda_app}
\end{align}
and
\begin{align}
m_2^{2[2]}=m_2^{2[1]}A_2^{-1/3},
\qquad
m_1^{2[2]}
=
m_1^{2[1]}
+\frac{\lambda_3^{[1]}m_2^{2[1]}}{\lambda_2^{[1]}}
\left(1-A_2^{1/3}\right).
\label{eq:stage2mass_app}
\end{align}
Finally,
\begin{align}
\lambda_3(t_{12})
=
\frac{\lambda_3^{[2]}}{1-\mathfrak b\,\lambda_3^{[2]}t_{12}},
\qquad
\lambda_1=\lambda_1^{[2]},
\qquad
\lambda_2=\lambda_2^{[2]},
\qquad
m_i^2=m_i^{2[2]}.
\label{eq:stage3_app}
\end{align}
The sequential structure is justified because the noncommutativity of distinct partial flows begins only at higher order. At one loop, one may therefore think of the multiscale RG trajectory as being built in three conceptually simple passes: first evolve the field-1 diagonal sector, then the field-2 diagonal sector, and only at the end let the mixed logarithm renormalize the portal coupling.

The optimized trajectory is defined by demanding that the explicit one-loop logarithms vanish:
\begin{align}
L_1(\varphi;\vec t^{\,*})=0,
\qquad
L_2(\varphi;\vec t^{\,*})=0,
\qquad
L_{12}(\varphi;\vec t^{\,*})=0.
\label{eq:optimal_logs_main}
\end{align}
In the positivity domain these conditions amount to
\begin{align}
\kappa_1^{2*}(\varphi)=\frac{M_1^2(\varphi;\vec t^{\,*})}{\bar u_1^2},
\qquad
\kappa_2^{2*}(\varphi)=\frac{M_2^2(\varphi;\vec t^{\,*})}{\bar u_2^2},
\label{eq:kappa12star_main}
\end{align}
and
\begin{align}
\ln \kappa_{12}^{2*}(\varphi)
=
\frac{1}{\mathcal J_{12}}
\int_0^1 dx\,\frac{1}{\Xi(x)}
\ln\!\left(\frac{\mathcal M(x;\varphi,\vec t^{\,*})^2}{\bar u(x)^2}\right),
\label{eq:kappa12star_mixed_main}
\end{align}
where $\mathcal M(x)^2=xM_1^2+(1-x)M_2^2$. The strict leading-log (LL) improved potential is then simply the tree-level expression evaluated on the optimized characteristic:
\begin{align}
V_{\rm LL}^{\rm multiscaleRG}(\varphi_1,\varphi_2)
=
\frac12 m_1^2(\vec t^{\,*})\varphi_1^2
+\frac12 m_2^2(\vec t^{\,*})\varphi_2^2
+\frac{\lambda_1(\vec t^{\,*})}{24}\varphi_1^4
+\frac{\lambda_2(\vec t^{\,*})}{24}\varphi_2^4
+\frac{\lambda_3(\vec t^{\,*})}{4}\varphi_1^2\varphi_2^2.
\label{eq:VLL_main}
\end{align}
This is the natural multiscale generalization of the usual RG-improved Coleman--Weinberg potential. From a practical point of view, the strategy is simple: first solve the three characteristic flows, then choose the point along that three-parameter trajectory at which all explicit logarithms disappear. What remains is a tree-level-looking potential whose couplings already remember which logarithmic sector produced them.

\section{Isotropic but unequal velocities}
\label{sec:isotropic}

A large amount of analytic control is recovered when the fully directional anisotropy is reduced to a single isotropic velocity for each field,
\begin{align}
\mC_1=u_1^2 I,
\qquad
\mC_2=u_2^2 I,
\qquad
u_1\neq u_2.
\label{eq:isotropicunequal_main}
\end{align}
Then $Q_i(\mathbf k)=u_i^2\mathbf k^2$ and the exact one-loop determinant reduces to a radial integral. More importantly, the mixed ultraviolet weight becomes elementary:
\begin{align}
\mathcal J_{12}
=
\int_0^1 dx\,\frac{1}{\big(xu_1^2+(1-x)u_2^2\big)^{3/2}}
=
\frac{2}{u_1u_2(u_1+u_2)}.
\label{eq:J12_iso_main}
\end{align}
The mixed logarithmic kernel also admits a closed form. Define
\begin{align}
\mathcal K(A,B;u_1,u_2;\mu)
&\equiv
\int_0^1 dx\,
\frac{1}{\big(xu_1^2+(1-x)u_2^2\big)^{3/2}}
\ln\!\left(\frac{xA+(1-x)B}{\mu^2\big(xu_1^2+(1-x)u_2^2\big)}\right),
\label{eq:K_iso_def_main}
\end{align}
where $A$ and $B$ are positive mass-squared arguments. The integrand contains two competing linear interpolations: one in the mass sector, $xA+(1-x)B$, and one in the velocity sector, $xu_1^2+(1-x)u_2^2$. A convenient change of variables is to trade the Feynman parameter $x$ for the interpolating speed itself. After this step the logarithm becomes a rational function of the new variable and the integral can be done in closed form. The result is
\begin{align}
\mathcal K(A,B;u_1,u_2;\mu)
=
\frac{2}{u_1^2-u_2^2}
\left[
\frac{2-\ln\!\left(\frac{A}{\mu^2u_1^2}\right)}{u_1}
-\frac{2-\ln\!\left(\frac{B}{\mu^2u_2^2}\right)}{u_2}
+\Phi(u_1)-\Phi(u_2)
\right],
\label{eq:K_iso_closed_main}
\end{align}
where
\begin{align}
\alpha\equiv \frac{A-B}{u_1^2-u_2^2},
\qquad
\beta\equiv \frac{u_1^2B-u_2^2A}{u_1^2-u_2^2},
\qquad
\chi\equiv \sqrt{\left|\frac{\alpha}{\beta}\right|},
\label{eq:alpha_beta_chi_main}
\end{align}
and
\begin{align}
\Phi(y)= \begin{cases}2 \chi \arctan (\chi y), & \alpha>0, \beta>0, \\ -2 \chi \operatorname{artanh}(\chi y), & \alpha<0, \beta>0, \\ -2 \chi \operatorname{arcoth}(\chi y), & \alpha>0, \beta<0 .\end{cases}
\label{eq:Phi_main}
\end{align}
The auxiliary function $\Phi$ is the compact way of keeping track of the analytic branches. When $\alpha/\beta>0$ the integral is trigonometric; when $\alpha/\beta<0$ it is hyperbolic; and the single symbol $\Phi$ keeps the final expression readable without hiding this analytic distinction.

The local effective potential reduces correspondingly to
\begin{align}
V_{\rm eff}^{\rm local}
&=
V_{\rm tree}
+\sum_{i=1}^{2}
\frac{1}{64\pi^2u_i^3}
\left[
2m_i^2U_i\left(\ln\!\frac{m_i^2}{\mu^2u_i^2}-1\right)
+U_i^2\ln\!\frac{m_i^2}{\mu^2u_i^2}
\right]
\nonumber\\
&\qquad
+\frac{s^2}{32\pi^2}\,
\mathcal K(m_1^2,m_2^2;u_1,u_2;\mu)
+\cO(\varphi^6),
\label{eq:Vlocal_iso_main}
\end{align}
and the quartic beta functions become
\begin{align}
\beta_{\lambda_1}
&=
\frac{3}{16\pi^2}
\left(
\frac{\lambda_1^2}{u_1^3}+\frac{\lambda_3^2}{u_2^3}
\right),
\nonumber\\
\beta_{\lambda_2}
&=
\frac{3}{16\pi^2}
\left(
\frac{\lambda_2^2}{u_2^3}+\frac{\lambda_3^2}{u_1^3}
\right),
\nonumber\\
\beta_{\lambda_3}
&=
\frac{\lambda_3}{16\pi^2}
\left(
\frac{\lambda_1}{u_1^3}
+\frac{\lambda_2}{u_2^3}
+\frac{8\lambda_3}{u_1u_2(u_1+u_2)}
\right).
\label{eq:betas_iso_main}
\end{align}
This specialization is also the cleanest place to see an accidental RG-invariant ray. Introduce the rescaled couplings
\begin{align}
x\equiv \frac{\lambda_1}{u_1^3},
\qquad
y\equiv \frac{\lambda_2}{u_2^3},
\qquad
z\equiv \frac{\lambda_3}{u_1^{3/2}u_2^{3/2}},
\qquad
\rho\equiv \frac{4\sqrt{u_1u_2}}{u_1+u_2}\le 2.
\label{eq:xyzrho_main}
\end{align}
Then the symmetric ray
\begin{align}
x=y=\xi z
\label{eq:ray_ansatz_main}
\end{align}
is preserved by the one-loop flow provided
\begin{align}
\xi^2-2\rho\,\xi+3=0.
\label{eq:kappa_eq_main}
\end{align}
Hence, whenever $\rho^2\ge 3$, there are real branches
\begin{align}
\xi=\rho\pm\sqrt{\rho^2-3},
\label{eq:kappa_solutions_main}
\end{align}
which define an accidental one-parameter fixed line in the space of coupling ratios. In the equal-velocity limit $\rho=2$, this reproduces the two familiar symmetric rays.

The multiscale matching problem also becomes fully explicit in this isotropic setting. In particular,
\begin{align}
\kappa_1^{2*}=\frac{A}{u_1^2},
\qquad
\kappa_2^{2*}=\frac{B}{u_2^2},
\label{eq:kappa12_iso_main}
\end{align}
while the mixed scale is determined by $\mathcal K(A,B;u_1,u_2;\kappa_{12}^*)=0$. Since $\mathcal K$ is linear in $\ln\mu^2$, one may write directly
\begin{align}
\ln\kappa_{12}^{2*}
=
\frac{\mathcal K(A,B;u_1,u_2;1)}{\mathcal J_{12}},
\label{eq:kappa12match_app}
\end{align}
that is,
\begin{align}
\ln\kappa_{12}^{2*}
=
\frac{u_1u_2(u_1+u_2)}{u_1^2-u_2^2}
\left[
\frac{2-\ln\!\left(\frac{A}{u_1^2}\right)}{u_1}
-\frac{2-\ln\!\left(\frac{B}{u_2^2}\right)}{u_2}
+\Phi(u_1)-\Phi(u_2)
\right].
\label{eq:kappa12closed_app}
\end{align}
When $A/u_1^2=B/u_2^2$, all three optimal scales coincide.

Several useful limits are worth isolating because each of them condenses a different aspect of the mixed logarithm. If $u_1=u_2=u$, the kernel reduces to
\begin{align}
\mathcal K(A,B;u,u;\mu)
=
\frac{1}{u^3}
\left[
\frac{A\ln\!\frac{A}{\mu^2u^2}-B\ln\!\frac{B}{\mu^2u^2}}{A-B}
-1
\right].
\label{eq:K_equalu_app}
\end{align}
If the masses and velocities are proportional,
\begin{align}
\frac{A}{u_1^2}=\frac{B}{u_2^2}\equiv c,
\label{eq:proportional_app}
\end{align}
then
\begin{align}
\mathcal K(A,B;u_1,u_2;\mu)
=
\mathcal J_{12}\ln\!\left(\frac{c}{\mu^2}\right)
=
\frac{2}{u_1u_2(u_1+u_2)}
\ln\!\left(\frac{A/u_1^2}{\mu^2}\right).
\label{eq:K_proportional_app}
\end{align}
For equal velocities the mixed matching scale becomes
\begin{align}
\kappa_{12}^{2*}
=
\frac{1}{eu^2}
\exp\!\left[
\frac{A\ln A-B\ln B}{A-B}
\right].
\label{eq:identric_app}
\end{align}
Finally, the higher mixed coefficients themselves admit a neat closed representation. Writing
\begin{align}
u_2^2(1-px)=xu_1^2+(1-x)u_2^2,
\qquad
B(1-qx)=xA+(1-x)B,
\label{eq:pq_defs_app}
\end{align}
with $p=1-u_1^2/u_2^2$ and $q=1-A/B$, one recognizes
\begin{align}
J_n(A,B;u_1,u_2)
=
\frac{1}{16\pi^2}
\frac{1}{(2n-1)(2n-2)\,u_2^3\,B^{2n-2}}
F_1\!\left(
n;\frac32,2n-2;2n;
1-\frac{u_1^2}{u_2^2},
1-\frac{A}{B}
\right),
\label{eq:Jn_Appell_app}
\end{align}
where $F_1$ is the Appell function. This formula is not needed for the later sections, but it makes clear that the entire higher mixed tower remains under analytic control in the isotropic unequal-velocity limit.

\section{Broken-phase dispersion relations in the local-potential approximation}
\label{sec:broken}

The full pole spectrum in a broken phase is controlled by the momentum-dependent two-point function. At tree level, or more generally within the local-potential approximation in which momentum-dependent self-energies are neglected, the spectrum is nevertheless fully explicit. We restrict ourselves to that setting here.

For isotropic microscopic kinetic terms, let $(v_1,v_2)$ be a homogeneous vacuum and write
\begin{align}
\phi_1=v_1+h_1,
\qquad
\phi_2=v_2+h_2.
\label{eq:vacuum_split_main}
\end{align}
Let us also define the following matrix of the effective potential at the vacuum is
\begin{align}
\mathbb M_v^2
=
\begin{pmatrix}
\mathcal A & \mathcal C\\
\mathcal C & \mathcal B
\end{pmatrix},
\label{eq:curvature_matrix_main}
\end{align}
with
\begin{align}
\mathcal A=\left.\frac{\partial^2V_{\rm eff}}{\partial\phi_1^2}\right|_v,
\qquad
\mathcal B=\left.\frac{\partial^2V_{\rm eff}}{\partial\phi_2^2}\right|_v,
\qquad
\mathcal C=\left.\frac{\partial^2V_{\rm eff}}{\partial\phi_1\partial\phi_2}\right|_v.
\label{eq:ABC_main}
\end{align}
At tree level,
\begin{align}
\mathcal A
=
m_1^2+\frac{\lambda_1}{2}v_1^2+\frac{\lambda_3}{2}v_2^2,
\qquad
\mathcal B
=
m_2^2+\frac{\lambda_2}{2}v_2^2+\frac{\lambda_3}{2}v_1^2,
\qquad
\mathcal C=\lambda_3v_1v_2.
\label{eq:ABC_tree_main}
\end{align}

For plane waves $h_i\sim e^{-iEt+i\mathbf p\cdot\mathbf x}$, the inverse propagator is
\begin{align}
\Gamma^{-1}(E,p)
=
\begin{pmatrix}
E^2-u_1^2p^2-\mathcal A & -\mathcal C\\
-\mathcal C & E^2-u_2^2p^2-\mathcal B
\end{pmatrix},
\qquad
p\equiv |\mathbf p|.
\label{eq:inverseprop_main}
\end{align}
The pole condition therefore gives the exact local-potential branches
\begin{align}
E_\pm^2(p)
=
\frac{
\mathcal A+\mathcal B+(u_1^2+u_2^2)p^2
\pm
\sqrt{\big[\mathcal A-\mathcal B+(u_1^2-u_2^2)p^2\big]^2+4\mathcal C^2}
}{2}.
\label{eq:dispersion_exact_main}
\end{align}
At zero momentum the masses are
\begin{align}
m_H^2=\frac{\mathcal A+\mathcal B+\Delta}{2},
\qquad
m_L^2=\frac{\mathcal A+\mathcal B-\Delta}{2},
\qquad
\Delta\equiv \sqrt{(\mathcal A-\mathcal B)^2+4\mathcal C^2}.
\label{eq:masses_zero_main}
\end{align}
Rotating to the mass basis diagonalizes the $\mathbb M_v^2$ matrix but, unless $u_1=u_2$, it does not diagonalize the spatial kinetic term. The matrix that diagonalizes one part of the dynamics does not diagonalize the full kinematics.

It is therefore useful to distinguish two angles. The first is the vacuum or mass-mixing angle, which diagonalizes the $\mathbb M_v^2$ matrix at $p=0$:
\begin{align}
\tan 2\theta_v = \frac{2\mathcal C}{\mathcal A-\mathcal B},
\qquad
\cos 2\theta_v=\frac{\mathcal A-\mathcal B}{\Delta},
\qquad
\sin 2\theta_v=\frac{2\mathcal C}{\Delta}.
\label{eq:thetav_app}
\end{align}
With the rotation
\begin{align}
\begin{pmatrix}
h_1\\ h_2
\end{pmatrix}
=
R(\theta_v)
\begin{pmatrix}
H\\ L
\end{pmatrix},
\qquad
R(\theta_v)=
\begin{pmatrix}
\cos\theta_v & -\sin\theta_v\\
\sin\theta_v & \cos\theta_v
\end{pmatrix},
\label{eq:Rtheta_app}
\end{align}
the mass matrix becomes diagonal, but the velocity matrix turns into
\begin{align}
R(\theta_v)^T
\begin{pmatrix}
u_1^2 & 0\\
0 & u_2^2
\end{pmatrix}
R(\theta_v)
=
\begin{pmatrix}
u_H^2 & y\\
y & u_L^2
\end{pmatrix},
\label{eq:velocitymatrix_app}
\end{align}
with
\begin{align}
u_H^2=u_1^2\cos^2\theta_v+u_2^2\sin^2\theta_v,
\qquad
u_L^2=u_1^2\sin^2\theta_v+u_2^2\cos^2\theta_v,
\qquad
y=\frac{u_2^2-u_1^2}{2}\sin 2\theta_v.
\label{eq:uHy_app}
\end{align}
The quantity $y$ is the off-diagonal entry of the velocity matrix in the mass basis. It vanishes only if the microscopic speeds are equal or the vacuum mixing angle is trivial. For nonzero $y$, the propagating modes at finite momentum are not the same as the zero-momentum mass eigenstates.

The corresponding momentum-dependent diagonalization angle in the mass basis is determined by
\begin{align}
\tan 2\vartheta(p)
=
\frac{2yp^2}{m_H^2-m_L^2+(u_H^2-u_L^2)p^2}
=
\frac{(u_2^2-u_1^2)\sin 2\theta_v\,p^2}
{\Delta+(u_1^2-u_2^2)\cos 2\theta_v\,p^2}.
\label{eq:thetavartheta_main}
\end{align}
This formula has a simple physical interpretation: as $p\to 0$, one has $\vartheta(p)\to 0$, so the propagating states coincide with the mass eigenstates; at large momentum the velocity matrix dominates, and the propagating states rotate back toward the original microscopic fields.

The low-momentum propagation speeds may be written directly as
\begin{align}
u_H^2
&=
\frac{u_1^2+u_2^2}{2}
+\frac{u_1^2-u_2^2}{2}\,
\frac{\mathcal A-\mathcal B}{\sqrt{(\mathcal A-\mathcal B)^2+4\mathcal C^2}},
\nonumber\\
u_L^2
&=
\frac{u_1^2+u_2^2}{2}
-\frac{u_1^2-u_2^2}{2}\,
\frac{\mathcal A-\mathcal B}{\sqrt{(\mathcal A-\mathcal B)^2+4\mathcal C^2}},
\label{eq:uHL_main}
\end{align}
while the pole condition in the mass basis takes the compact form
\begin{align}
(E^2-u_H^2p^2-m_H^2)(E^2-u_L^2p^2-m_L^2)-y^2p^4=0.
\label{eq:pole_massbasis_app}
\end{align}
The small-momentum expansion then reads
\begin{align}
E_H^2(p)
&=
m_H^2+u_H^2p^2
+
\frac{y^2}{m_H^2-m_L^2}\,p^4
+\cO(p^6),
\nonumber\\
E_L^2(p)
&=
m_L^2+u_L^2p^2
-
\frac{y^2}{m_H^2-m_L^2}\,p^4
+\cO(p^6).
\label{eq:smallp_massbasis_app}
\end{align}
Using \eqref{eq:thetav_app} and \eqref{eq:uHy_app}, one recovers the equivalent form
\begin{align}
E_H^2(p)
&=
m_H^2+u_H^2p^2
+
\frac{(u_1^2-u_2^2)^2\mathcal C^2}{\big[(\mathcal A-\mathcal B)^2+4\mathcal C^2\big]^{3/2}}\,p^4
+\cO(p^6),
\nonumber\\
E_L^2(p)
&=
m_L^2+u_L^2p^2
-
\frac{(u_1^2-u_2^2)^2\mathcal C^2}{\big[(\mathcal A-\mathcal B)^2+4\mathcal C^2\big]^{3/2}}\,p^4
+\cO(p^6),
\label{eq:smallp_main}
\end{align}
quoted earlier in compact form. The $p^4$ terms are therefore not mysterious higher-derivative corrections; they are simply the first residual effect of diagonalizing the mass matrix and the velocity matrix with different rotations. They vanish if either the microscopic speeds coincide or the off-diagonal $\mathcal C$ vanishes. At large momentum, by contrast, the propagating modes resolve the microscopic fields and recover the original velocities $u_1$ and $u_2$.

For the genuinely mixed broken phase with $v_1v_2\neq 0$, the tree-level stationarity equations are
\begin{align}
m_1^2+\frac{\lambda_1}{6}v_1^2+\frac{\lambda_3}{2}v_2^2=0,
\qquad
m_2^2+\frac{\lambda_2}{6}v_2^2+\frac{\lambda_3}{2}v_1^2=0.
\label{eq:stationarity_mixed_app}
\end{align}
Solving gives
\begin{align}
v_1^2
=
\frac{6(3\lambda_3m_2^2-\lambda_2m_1^2)}{\lambda_1\lambda_2-9\lambda_3^2},
\qquad
v_2^2
=
\frac{6(3\lambda_3m_1^2-\lambda_1m_2^2)}{\lambda_1\lambda_2-9\lambda_3^2}.
\label{eq:v1v2_app}
\end{align}
Substituting these back into the $\mathcal A, \mathcal{B}, \mathcal{C}$ yields
\begin{align}
\mathcal A=\frac{\lambda_1}{3}v_1^2,
\qquad
\mathcal B=\frac{\lambda_2}{3}v_2^2,
\qquad
\mathcal C=\lambda_3v_1v_2,
\label{eq:ABC_mixed_main}
\end{align}
so the exact local-potential dispersion relation follows from \eqref{eq:dispersion_exact_main} by direct substitution.

Along the Gildener--Weinberg ray the light state is the scalon; at tree level $m_{\rm scalon}^2=0$, while at one loop
\begin{align}
m_{\rm scalon}^2=8Bv^2,
\label{eq:scalon_broken_main}
\end{align}
with $B$ given in \eqref{eq:B_main}. The low-energy velocities simplify to
\begin{align}
u_{\rm scalon}^2=u_1^2n_1^2+u_2^2n_2^2,
\qquad
u_H^2=u_1^2n_2^2+u_2^2n_1^2.
\label{eq:GW_velocities_main}
\end{align}
The full one-loop pole spectrum would require the momentum-dependent self-energy matrix and lies beyond the local-potential approximation adopted here.

\section{Conclusions and outlook}

We studied scalar quantum field theories in which different fields propagate with inequivalent spatial kinetic tensors, i.e.\ with genuinely distinct propagation cones.  The main message is that this ``disparity in sound speeds'' is not a cosmetic deformation of an isotropic problem: it changes the kinematic object on which elastic unitarity acts and it imprints a structured set of geometric invariants on the ultraviolet sector of the effective action.

On the scattering side, we derived an exact elastic unitarity relation in the center-of-mass frame in which the two-body phase space becomes a positive \textit{kernel} on $S^2$. The familiar partial-wave amplitudes are promoted to matrices in angular space, and the sharp elastic bound constrains the eigenvalues of the rescaled operator $G^{1/2}aG^{1/2}$ rather than individual entries of the channel amplitude. In the weakly anisotropic regime the phase-space kernel contains only monopole and quadrupole moments at first order, so the first nontrivial operator-level effect is the controlled mixing of $s$- and $d$-waves.

In the renormalizable two-scalar quartic model, this structure is visible already at one loop: although the tree amplitude is angle independent, the bubble discontinuity is weighted by the anisotropic on-shell measure.  The explicit absorptive-part computation reproduces the general optical-theorem form and makes transparent which aspects of the dynamics are universal (threshold exponents, positivity) and which are geometric (directional weights and their angular harmonics). 

On the effective-action side, we showed that the mismatch of the spatial tensors persists in the one-loop determinant and organizes the local renormalizable sector into distinct geometric structures.  Diagonal contractions are controlled by $\Pi_i^{-1}=1/\sqrt{\det \mC_i}$, while the genuinely mixed sector is governed by the interpolating invariant
$\mathcal J_{12}=\int_0^1 dx\,(\det[x\mC_1+(1-x)\mC_2])^{-1/2}$.
In the classically scale-invariant regime the flat direction remains the standard Gildener--Weinberg ray, but the one-loop coefficient determining the scalon mass acquires separate diagonal and mixed geometric weights.  The resulting three-log structure is naturally handled by a three-scale multiscale RG improvement.

This whole structure is consistent with the broader renormalization pattern found in the more general anisotropic two-scalar theory of Ref.~\cite{Ageev:2025bwc}, where the one-loop UV data admit a basis-covariant organization in terms of anisotropy-dependent geometric kernels.

There are several direct extensions of this work.  First, it would be natural to generalize the exact unitarity framework to interactions with \textit{derivatives} (or nonlinear sigma-model kinetic terms), where angular dependence enters not only through phase space but also through the vertices.  Second, while the present quartic model has no one-loop wavefunction renormalization, at higher loop order—and in theories with Yukawa or gauge couplings—one expects momentum-dependent self-energies and hence running of the Lorentz-violating tensors, connecting directly to the broader renormalization literature on Lorentz-violating scalar, Yukawa, and gauge theories \cite{Altschul:2006Yukawa,Ferrero:2011RenLV,Altschul:2012ScalarPot,Altschul:2022ScalarQED,Altschul:2023ScalarQCD}. Third, it would be interesting to explore UV completions in which disparate sound cones emerge after integrating out a heavy field-space direction, producing an effective single-field description with nontrivial kinetic structure for a gapless mode \cite{Babichev:2018twg}.  Finally, since superluminality and singular sound cones often signal a tension with analyticity and UV completion, it would be valuable to combine the operator unitarity bounds developed here with dispersive/analytic constraints in the spirit of \cite{Adams:2006sv}, now in genuinely multi-cone settings.

\section*{Acknowledgments}

The authors are grateful to Irina Aref’eva and Dmitry Gorbunov for useful comments and
fruitful discussions. This work has been supported by Russian Science Foundation Grant
No. 24-72-00121, https://rscf.ru/project/24-72-00121/.

\end{document}